\documentclass[a4paper,12pt]{article}
\pdfoutput=1
\usepackage[T1]{fontenc}

\usepackage{comment}
\usepackage{jcappub}
\usepackage{subcaption}
\usepackage{hyperref}
\usepackage{graphicx}
\usepackage{xcolor}
\usepackage{amssymb,amsmath,bm}
\usepackage[utf8]{inputenc}
\usepackage[export]{adjustbox}
\hypersetup{colorlinks=true, citecolor=blue}
\usepackage{multirow}
\usepackage{subcaption}
\usepackage{amsmath, amssymb, amsfonts}
\usepackage{siunitx}
\usepackage{booktabs}
\usepackage{mathtools}
\usepackage{placeins}
\usepackage{anyfontsize}
\def \bn {\mathbf{n}}

\newcommand{\vb}[1]{\mathbf{#1}}
\usepackage[outercaption]{sidecap}
\usepackage{subfiles}
\LetLtxMacro{\originaleqref}{\eqref}
\renewcommand{\eqref}{Eq.~\originaleqref}

\usepackage[toc, page]{appendix}
\usepackage{tikz}
\usetikzlibrary{calc, intersections, through, arrows, backgrounds}
\usetikzlibrary{arrows.meta}

\hypersetup{
    colorlinks=true,
    citecolor=blue,
    linkcolor=blue,
    linktoc=page
}

\title{\fontsize{20}{32}\selectfont{The power spectrum of luminosity distance fluctuations in General Relativity \vspace{-0.2in} \\}}

\author[a]{\fontsize{15}{25}\selectfont Mattia Pantiri,}
\author[b]{Matteo Foglieni,}
\author[c]{Enea Di Dio,}
\author[d,e]{Emanuele Castorina}

\affiliation[a]{Institute Lorentz, Leiden University, PO Box 9506, Leiden 2300 RA, The Netherlands}
\affiliation[b]{Leibniz Supercomputing Centre (LRZ), Boltzmannstraße 1, 85748 Garching bei München, Germany}
\affiliation[c]{D\'epartement de Physique Th\'eorique and Center for Astroparticle Physics,
Universit\'e de Gen\`eve, Quai E.~Ansermet 24, CH-1211 Gen\`eve 4, Switzerland}
\affiliation[d]{Dipartimento di Fisica ‘Aldo Pontremoli’, Universita’ degli Studi di Milano, \\ Via Celoria 16, 20133 Milan, Italy}
\affiliation[e]{INFN, Sezione di Milano, Via Celoria 16, 20133 Milan, Italy}

\emailAdd{\textcolor{blue}{pantiri@lorentz.leidenuniv.nl}}
\emailAdd{\textcolor{blue}{matteo.foglieni@lrz.de}}
\emailAdd{\textcolor{blue}{enea.didio@unige.ch}}
\emailAdd{\textcolor{blue}{emanuele.castorina@unimi.it}}

\abstract{At low redshift, it is possible to combine spectroscopic information of galaxies with their luminosity or angular diameter distance to directly measure the projection of peculiar velocities (PV) along the line-of-sight. A PV survey probing a large fraction of the sky is subject to so-called wide-angle effects, arising from the variation of the line-of-sight across the sky, and other sub-leading projection effects due to the propagation of the photons in a perturbed cosmological background. In this work, for the first time, we provide a complete description, within linear theory and General Relativity, of the power spectrum of luminosity distance fluctuations, clarifying its relation to the observables in a PV survey. We find that wide-angle effects will be detected at high significance by future observations and will have to be included in the cosmological analysis. Other relativistic projections effects could also be detected provided accurate, per object, distances are available.}

\newcommand{\dd}{\mathrm{d}}

\newcommand{\Jld}{\mathcal{J}}
\newcommand{\Jcr}{\mathfrak{J}}

\newcommand{\HH}{{\cal H}}
\newcommand{\HHd}{\dot{\cal H}}
\newcommand{\mb}{s_\mathrm{b}}
\newcommand{\mbs}[1]{s_{\mathrm{b}\ifx &#1&\else ,#1\fi}}
\newcommand{\fevo}{f_\mathrm{evo}}
\newcommand{\fevos}[1]{f_{\mathrm{evo}, #1}}

\newcommand{\Rld}{\mathfrak{R}}
\newcommand{\Rgnc}{\mathcal{R}}

\newcommand{\omegaMo}{\Omega_{\mathrm{M}0}}

\newcommand{\ndv}{{v_{\parallel}}}

\newcommand{\vs}{\mathbf{s}}
\newcommand{\vk}{\mathbf{k}}

\newcommand{\versor}[1]{\hat{\mathbf{#1}}}

\newcommand{\Legendre}[2]{\mathcal{L}_{#1}\left( #2 \right)}

\newcommand{\avg}[1]{\left\langle #1 \right\rangle}

\newcommand{\tj}[6]{ \begin{pmatrix}
   #1 & #2 & #3 \\
   #4 & #5 & #6
  \end{pmatrix}}

\begin{document}

\maketitle

\section{Introduction}
The motion of objects in an expanding Universe contains a wealth of information about the growth of the Large Scale Structure (LSS). Traditionally, the peculiar velocity field has been studied through its imprint on the clustering of galaxies, an effect known as Redshift Space Distortions (RSD) \cite{Kaiser_1987}. The measured redshift of an object indeed contains a term proportional to its peculiar velocity, producing a distinct anisotropic signature in the large scale distribution of galaxies that can be used to probe the standard cosmological model and its extensions \cite{weinberg13}.

Alternatively, the peculiar motion of a galaxy, or better said its projection along the line-of-sight (LOS), can be inferred from its redshift if another distance measurement of the object is available. The latter could be obtained via empirical relations, \emph{e.g.} The Tully-Fisher relation \cite{TullyFisher} or the Fundamental Plane~\cite{FunPlane,Dressles1987}, or by directly observing a standard candle like a Supernova Type-1A inside the galaxy. A collection of peculiar motions for several galaxies is then called a peculiar velocity (PV) surveys~\cite{Park:2000rc}.  These surveys could measure the growth of structure at the \% level~\cite{Koda:2013eya,howlett2017cosmological}, and provide constraints on neutrino masses and modification of gravity competitive with traditional galaxy redshfit surveys \cite{Gordon:2007zw,Amendola:2019lvy,Johnson:2015aaa,howlett2017cosmological,Quartin:2021dmr,whitford2022using}.
PV surveys have been carried out within 2MTF \cite{Howlett:2017asq,Qin:2019axr}, 6dF \cite{Adams:2017val,Adams:2020dzw,Qin:2019axr,Johnson:2014kaa} and SDSS/BOSS \cite{Howlett:2022len} data, and
will see an order of magnitude improvement in their statistical accuracy within the coming decade \cite{Howlett:2017asw,DESIPV,kim2019testing}.
Next generation gravitational wave observatories could also provide peculiar velocity information for a large number of sources and obtain competitive cosmological constraints \cite{Palmese:2020kxn}.

As the uncertainties of the measurements become smaller and smaller, the modeling of the peculiar velocities will need to be improved accordingly.
In particular, accurate redshift-independent distances of galaxies can be obtained only at low-redshift, which in turns requires to cover large areas of the sky to access a larger volume.
The combination of the two means that, for PV surveys, the distant observer approximation, or plane-parallel limit, breaks down and it is important to model the effect of a varying LOS across the sky. These so-called wide-angle effects can lead to several tens of percent deviation from the plane-parallel limit at low-redshift, and have to be modeled in both the auto-power spectrum of peculiar velocities, as well as in their cross correlation with the galaxy density field \cite{Nusser:2017qdc,Gorski88,Castorina_2019,Lai:2022sgp,Adams:2020dzw}.

At the same time, distance indicators like the TF relation, the FP or SN1A, truly measure the luminosity distance to a given object, and thus provide us not just with their peculiar velocities, but with additional distortions \footnote{The Tully-Fisher relation actually measures fluctuations in the angular diameter distance. The latter is related to luminosity distances via Etherington's reciprocity theorem \cite{etherington1933lx}.}. The fluctuations of the luminosity distances in General Relativity (GR) actually contain several contributions other than the radial peculiar velocity \cite{Bonvin_2006}, and these could either bias the estimate of cosmological parameters, if not properly accounted for, or provide additional cosmological information.

The goal of this work is to compute, for the first time, both wide angle and GR corrections to the peculiar velocity 3-dimensional power spectra, also including the observational effects like the presence of a window function. The outline of this paper is as follows. In Sec.~\ref{sec:Lum} we review the formalism to describe the fluctuations of the luminosity distance in General Relativity. In Sec.~\ref{sec:estimator}
we present the estimator of the peculiar velocity power spectrum, compute its ensemble average and discuss a toy model for the survey window function. In Sec.~\ref{sec:results} we show the importance of wide angle and GR effects for a few representative examples, and forecast their detectability in Sec.~\ref{sec:forecasts}. We then conclude in Sec.~\ref{sec:conclude}.
All the numerical calculations presented in this work have been obtained using our publicly available code GaPSE \footnote{\href{https://github.com/foglienimatteo/GaPSE.jl}{https://github.com/foglienimatteo/GaPSE.jl}}.

\section{The luminosity distance in General Relativity}
\label{sec:Lum}
\begin{figure}
\centering
        \begin{tikzpicture}[scale=1,
            background rectangle/.style={fill=white},
            show background rectangle
                            ]

            \def\sdue{5.5}
            \def\stre{6.5}
            \def\mispunto{1pt}

            \def\xpuntouno{0.0}
            \def\ypuntouno{0.0}
            \node [name=O] at (\xpuntouno,\ypuntouno) {};

            \def\alphadue{30}
            \def\cosalphadue{ 0.8660254037844387 }
            \def\sinalphadue{ 0.5 }
            \def\xpuntodue{\xpuntouno + \sdue * \cosalphadue}
            \def\ypuntodue{\sdue * \sinalphadue}
            \node [name=A] at (\xpuntodue,\ypuntodue) {};

            \def\alphatre{80}
            \def\cosalphatre{ 0.17364817766693041 }
            \def\sinalphatre{ 0.984807753012208 }
            \def\xpuntotre{\xpuntouno + \stre * \cosalphatre}
            \def\ypuntotre{\stre * \sinalphatre}
            \node [name=B] at (\xpuntotre,\ypuntotre) {};

            \def\radciclephi{1.0}
            \def\angphi{105}
            \draw[line width = 0.3 mm, <->] (\xpuntodue + \cosalphadue * \radciclephi, \ypuntodue +  \sinalphadue * \radciclephi) arc (\alphadue:\angphi+\alphadue:\radciclephi);
            \draw[above] node at  (\xpuntodue + \cosalphadue  * \radciclephi - 0.5, 0.7 + \ypuntodue + \sinalphadue * \radciclephi) {$\varphi$};

            \def\radcicletheta{1.5}
            \draw[line width = 0.3 mm, <->] (\xpuntouno + \cosalphadue * \radcicletheta, \sinalphadue * \radcicletheta) arc (\alphadue:\alphatre:\radcicletheta);
            \draw[above] node at  (\xpuntouno + \cosalphadue  * \radcicletheta - 0.3, 0.5 + \sinalphadue * \radcicletheta) {$\theta$};

            \draw[below] node at (O) {$\mathbf{O}$};
            \draw[below right] node at (A) {$\mathbf{s}_1$};
            \draw[left] node at (B) {$\mathbf{s}_2$};
            \draw[above] node at (\xpuntotre * 0.5 + \xpuntodue * 0.5 , 0.76 * \ypuntotre) {$\mathbf{s}$};

            \filldraw[black, above]
                (O) circle (\mispunto)
                (A) circle (\mispunto)
                (B) circle (\mispunto);

            \draw[->, -triangle 90] (O) to (A);
            \draw[->, -triangle 90] (O) to (B);
            \draw[->, -triangle 90] (A) to (B);
            \draw[line width = 0.3mm, dotted] (A) to (\xpuntouno + 1.3*\sdue * \cosalphadue, 1.3*\sdue * \sinalphadue);

        \end{tikzpicture}
        \captionof{figure}{ Positions of the observer $\mathbf{O}$ and of the galaxies $\vs_1$ and $\vs_2$, together with their separation $\vs = \vs_2 - \vs_1$. }
        \label{fig:s1_s2_angles}
        \end{figure}
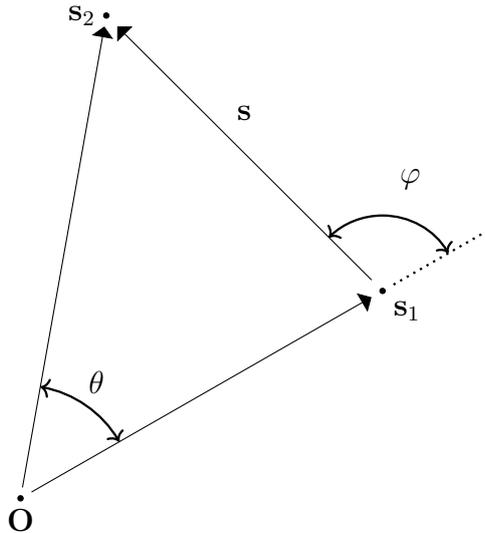

The fundamental observables of a peculiar velocity survey are fluctuations in the luminosity distance $D_L$.
In General Relativity, the expression for the linear luminosity distance perturbations is well known \cite{Bonvin_2006,biern2017gauge,Yoo:2016vne,Scaccabarozzi:2018vux,DiDio_2016}, and it reads
\begin{equation}
\label{eq:fluctuations_ld}
\begin{split}
    \Delta D_L =\frac{D_L - \langle D_L \rangle }{\langle D_L \rangle}
    =& -\bigg(\mathcal{H}_0V_0 + \Psi - \Psi_0 + v_{\parallel} - v_{\parallel_0} +  \int_0^r(\dot{\Phi} + \dot{\Psi})d\chi \bigg)\bigg(1 - \frac{1}{\mathcal{H}r}\bigg) + \\
    & - \int_0^r \frac{r - \chi}{r\chi}\Delta_{\Omega}\bigg(\frac{\Phi + \Psi}{2}\bigg)d\chi -\Phi - \frac{1}{r}V_0 - v_{\parallel_0} + \frac{1}{r}\int_0^r(\Phi + \Psi)d\chi\,.
\end{split}
\end{equation}
where $<..>$ denotes the angular average\footnote{See Appendix~\ref{app:mean_distance} for possible extra contributions to the perturbed luminosity distance arising from a different definition of the mean $\avg{D_L}$.} at fixed observed redshift, $v_\parallel = \mathbf{v}\cdot\mathbf{n}$ is the projection of the peculiar velocity $\mathbf{v}$ along the line of sight, where $\mathbf{n}$ is the unit vector pointing from the source to the observer, $V$ denotes the velocity potential and $\Phi$ and $\Psi$ are the Bardeen potentials\footnote{The luminosity distance as a function of redshift is an observable and, therefore, gauge invariant. Without loss of generality, we choose to express it in the Newtonian Gauge \begin{equation}
    ds^2 = a^2(\tau)[-(1+2\Psi)d\tau^2 + (1-2\Phi)d\mathbf{x}^2]\,,
\end{equation}
where the scalar metric perturbations reduce to the gauge-invariant Bardeen potentials.}. Quantities with a subscript $()_0$ are evaluated at the observer's position. The correlation function is then defined as
\begin{equation}
\label{eq:corr_function}
    \xi(s_1, s_2, \hat{\mathbf{s}}_1\cdot\hat{\mathbf{s}}_2 \equiv \cos{\theta}) =
        \bigg\langle \Delta D_L(\mathbf{s}_1) \Delta D_L(\mathbf{s}_2) \bigg\rangle\,,
\end{equation}
so that every contribution in \eqref{eq:fluctuations_ld} needs to be computed in auto correlation with itself and in cross correlation with the others. Being a function of the full triangle (see Fig.~\ref{fig:s1_s2_angles}) connecting the observer with the two sources, \eqref{eq:corr_function} is valid in full-sky.
Explicit expressions for all the terms in the auto-correlations and for the cross-correlation function between $\Delta{D_L}$ and galaxy number counts can be found in Appendix~\ref{APP_LD} and~\ref{APP_GNCxLD}.

We will often be interested in the multipoles of the correlation function expanded around the angle between the relative separation of the two galaxies $\mathbf{s}\equiv \mathbf{s}_2-\mathbf{s}_1$, and the line of sight to one of the two galaxies, for example $\mathbf{s}_1$. We then write
\begin{align}
\label{eq:xi_vv}
          \bigg\langle \Delta D_L (\mathbf{s}_1) \Delta D_L(\mathbf{s}_2) \bigg\rangle
     = \sum_\ell \xi_\ell (s,s_1)\mathcal{L}_\ell(\hat{\mathbf{s}}\cdot \hat{\mathbf{s}}_1)\,.
\end{align}

A graphic representation of the relevant geometric variables is presented in Fig.~\ref{fig:s1_s2_angles}. Since the two objects and the observer must form a triangle, we have the freedom to choose any set of three variables to describe the correlation function.
In the coordinate system used above, the correlation function turns out to be a linear combination of the following integrals
\begin{equation}
\label{Inl}
    I^n_\ell(s) = \int \frac{dq}{2\pi^2} q^2P(q) \frac{j_\ell(qs)}{(qs)^n}\,,
\end{equation}
where $P(q)$ is the linear matter power spectrum at $z=0$ and $j_\ell$ is the $\ell$-th spherical Bessel function.

\subsection{On the cancellation of IR divergences}\label{sec:infrared_divergence}
As we will see in Sec.~\ref{sec:estimator}, the calculation of the luminosity distance power spectrum is the convolution of the correlation function in \eqref{eq:xi_vv} with the multipoles of the survey window function. Since the expression for the luminosity distance describes an observable quantity, every statistical function related to it should be finite. However, a problem arises in such computation. Indeed, every individual term in the correlation function including the $I^4_0$ integral, \emph{e.g.}~two powers of the gravitational potentials, is divergent in the $q\to0$ limit, \emph{i.e.}~the infrared (IR) regime.
In this subsection we will show that even if the divergences appear in the individual terms of the correlation function, they exactly cancel each other when summed together to give the full correlation function.
The presentation here follows closely the one in \cite{Casto_DiDio} for galaxy number counts, and we will show that the IR cancellation in the luminosity distance fluctuations holds under the same conditions of the one for the galaxy number counts\footnote{Formally, the proof of the absence of IR divergences in the galaxy number counts implies the absence of IR divergences in the luminosity distance fluctuations. Indeed the terms proportional to the magnification bias parameter $s_b$ in the galaxy number counts can be written as
$$
\Delta_g \left( \bn , z \right) \supset 5 s_b \Delta D_L \, .
$$
}.
We begin by isolating the part of the correlation function which is actually divergent: consider the expression of the $I^4_0$ integral and add and subtract one from the spherical Bessel function
\begin{equation}
\label{add_and_subtract_1}
    I^4_0(s) = \int \frac{dq}{2\pi^2}q^2P(q)\frac{j_0(qs)}{(qs)^4} = \int \frac{dq}{2\pi^2}q^2P(q)\frac{j_0(qs) - 1 + 1}{(qs)^4} = \tilde{I}_0^4(s) + \int \frac{dq}{2\pi^2}q^2\frac{P(q)}{(qs)^4}\,,
\end{equation}
where we have defined the IR-safe quantity
\begin{equation}
    \tilde{I}^4_0(s) = \int \frac{dq}{2\pi^2}q^2P(q)\frac{j_0(qs) - 1}{(qs)^4}\,.
\end{equation}
Subtracting unity from the spherical Bessel function is equivalent to removing the variance (the contact term) from the correlation function.
We then formally rewrite the full correlation function as
\begin{equation}
    \xi = \sum \tilde{\xi} + \sum \sigma^2\,,
\end{equation}
where the sums run over all possible operators appearing \eqref{eq:fluctuations_ld}, $\tilde{\xi}$ is obtained with the replacement of all the $I^4_0$ with $\tilde{I}^4_0$ and the divergent part is relegated to the variance $\sigma^2$. The cancellation of IR divergences follows automatically if and only if
\begin{equation}
\label{eq:variance_cancellation}
    \sum \sigma^2 = 0\,.
\end{equation}
The goal of this Section is to discuss under what assumptions the condition above holds. It is easy to see that \eqref{eq:variance_cancellation} can be written as
\begin{equation}
\label{eq:cancellation}
    - \mathfrak{R}\bigg(\mathcal{H}_0 T_V(0) + T_\Psi(r) - T_\Psi(0) + \int_0^r (T_{\dot{\Psi}} + T_{\dot{\Phi}}) d\chi \bigg) - T_\Phi(r) - \frac{T_V(0)}{r}+ \frac{1}{r} \int_0^r (T_\Psi + T_\Phi) d\chi = 0\,,
\end{equation}
in terms of the linear theory, Fourier space transfer functions $T_\mathcal{O}(k,z)$, defined via the initial curvature perturbation $\zeta(k)$ as  $\mathcal{O} (k,z)
= \zeta(k) T_\mathcal{O}(k,z)$, and we introduced  $\mathfrak{R} = 1 - \frac{1}{\mathcal{H}r}$.
The first integral can be easily computed
\begin{equation}
\label{eq:integral1}
    \int_0^r (T_{\dot{\Psi}} + T_{\dot{\Phi}}) d\chi = T_\Psi(0) + T_\Phi(0) - T_\Psi(r) - T_\Phi(r)\,,
\end{equation}
while the second one needs a bit more work. First of all, in the absence of large scales anisotropic stress one has $\Psi = \Phi$, then, combining the Euler equation
\begin{equation}
    \dot{T}_V(r) + \mathcal{H}T_V(r) = T_\Phi(r)
\end{equation}
and the Weinberg condition for the existence of adiabatic modes \cite{weinberg2003adiabatic}
\begin{equation}
    T_\Phi(r) + \mathcal{H}T_V(r) = T_\Phi(0) + \mathcal{H}_0T_V(0)\,,
\end{equation}
we get the following expression for the transfer function of the gravitational potential
\begin{equation}
    T_\Phi(r) = \frac{1}{2}(\dot{T}_V(r) + T_\Phi(0) + \mathcal{H}_0T_V(0))\,.
\end{equation}
Recall that the Weinberg condition is valid in presence of adiabatic initial conditions, absence of large scales anisotropic stress, absence of local Primordial non Gaussianities and if General Relativity holds. Now also the second integral of \eqref{eq:cancellation} can be easily computed
\begin{equation}
\label{eq:integral2}
    \frac{1}{r} \int_0^r (T_\Psi + T_\Phi) d\chi = \frac{2}{r} \int_0^r T_\Phi d\chi = \frac{1}{r}(T_V(0) - T_V(r)) + T_\Phi(0) + \mathcal{H}_0T_V(0)\,.
\end{equation}
Substituting \eqref{eq:integral1} and \eqref{eq:integral2} in \eqref{eq:cancellation} we obtain
\begin{equation}
    T_\Phi(r) + \mathcal{H}T_V(r) = T_\Phi(0) + \mathcal{H}_0T_V(0)\,,
\end{equation}
which is again the Weinberg condition. Summing up, the IR divergences in the correlation function cancel out if and only if the Weinberg condition is satisfied, and we need it two times: firstly we need it in order to integrate the transfer function of $\Phi$, then in the final rewriting of \eqref{eq:variance_cancellation}.

In the correlation function, or the power spectrum, we have also terms which are proportional to one power of the metric potential. While these terms are not IR divergent in $\Lambda$CDM, we do expect the observable to be insensitive to these contributions.
Following the derivation of Ref.~\cite{Casto_DiDio}, one can indeed show that summing up all the relativistic contributions we find again an exact cancellation.

\section{The power spectrum estimator}
\label{sec:estimator}
An efficient compression of the cosmological information of peculiar velocity surveys is represented by the multipoles of the luminosity distance power spectrum. Following~\cite{Howlett:2019bky}, we define the estimator of multipoles $L$ of the luminosity distance power spectrum as
\begin{eqnarray}\label{eq:Yamamoto_estimator}
    \hat{P}_L (k) &=&
        \frac{2 L + 1}{A} \!
        \int\!\frac{\dd \Omega_\mathbf{k}}{4 \pi}
        \!\int\!\dd^3 \vs_1  \!\int\!\dd^3 \vs_2 \,
        \Delta{D_L}(\vs_1) \Delta{D_L}(\vs_2)
        \nonumber \\
        && \hspace{4cm}
        W(\vs_1) W(\vs_2)
        e^{i \vk \cdot ( \vs_1 - \vs_2) }
        \Legendre{L}{\versor{k} \cdot \versor{d}_{\rm LOS}}
        \,,
\end{eqnarray}
where $A$ is a normalization constant, $W(\mathbf{s})$ is the survey window function defining the angular footprint and redshift distribution of the data, $\mathcal{L}_L$ are Legendre polynomials, and $\hat{\mathbf{d}}_{\rm LOS}$ is the LOS direction we use to project the data. This estimator is completely analogous to the one used in galaxy redshift survey \cite{Yamamoto_2006} and can be further generalized to measure the cross-correlation between galaxies velocities and densities. As customary in power spectrum analyses, we will further assume $\mathbf{d}_{\rm LOS} = \vs_1$, which allows to estimate the power spectrum Fast Fourier Transforms (FFT)~\cite{Scoccimarro:2015bla,Bianchi:2015oia,Hand:2017irw}.

The measurements should then be compared to the analytical prediction for the ensemble average of the estimator in \eqref{eq:Yamamoto_estimator} which we write, following \cite{Casto_DiDio,Castorina:2017inr}, as

\begin{align}\label{eq:estimator_standard}
    \avg{\hat{P}_L(k)} &= \frac{2 L + 1}{A} \sum_{\ell=0}^{\infty}
        \sum_{\ell_1=0}^{\infty} (-i)^L
        \tj{L}{\ell}{\ell_1}{0}{0}{0}^2 (2 \ell_1 + 1)
        \nonumber \times \\[8pt]
        & \qquad\qquad
        \int\!\dd s \, s^2 \int\!\dd s_1 \, s_1^2 \,
        \xi_\ell(s, s_1)  \, j_L(k s) \, F_{\ell_1}( s_1 , s) \, .
\end{align}
where we have switched integration variables $\dd^3 \mathbf{s}_1 \dd^3 \mathbf{s}_2 \rightarrow \dd^3\mathbf{s}_1 \dd^3 \mathbf{s}$, and we formally defined the multipoles of the correlation function of the window function

\begin{equation}
\label{eq:Fell_def}
    F_{\ell_1} (s_1 , s) =
    \int \dd\Omega_{\versor{s}} \, \dd\Omega_{\versor{s}_1} \,
    W(\vs_1)  \, W(\vs_2) \,
    \Legendre{\ell_1}{\versor{s} \cdot \versor{s}_1}  \,.
\end{equation}
Completely analogous expressions can be found for the cross-power spectrum between galaxy number counts and luminosity distance fluctuations (in this work we assume the angular window function for both tracers is the same).

The explicit dependence on $s_1$ in the  equations above keeps track of any redshift evolution of the correlation function. It is however often assumed that one can evaluate the theoretical model at some effective redshift, $z_{\rm eff}$, defined by the selection function.
In this case, the expression in \eqref{eq:estimator_standard} further simplifies
\begin{equation}\label{eq:estimator_effective}
    \avg{\hat{P}_L(k)} = \frac{2 L + 1}{A} (-i)^L
    \sum_{\ell=0}^{\infty}
    \sum_{\ell_1=0}^{\infty}
    \tj{L}{\ell}{\ell_1}{0}{0}{0}^2
    \int\!\dd s \, s^2 \, \xi_\ell(s, s_{\rm eff}) \,
    j_L(k s) \, Q_{\ell_1}(s) \, ,
\end{equation}

\noindent
where we have defined the window functions multipole moments
\begin{equation}\label{eq:Q_ell}
    Q_{\ell_1}(s) = \int\!\dd s_1 \, s_1^2 \,
    F_{\ell_1}(s_1, s)\,,
\end{equation}
which can be estimated using standard pair-counting algoritmh \cite{Wilson:2015lup} or FFTs \cite{Beutler:2018vpe}.
Given the narrow extent in redshift of peculiar velocity surveys, with redshift bins of size $\Delta z \lesssim 0.2$, we expect the effective redshift approximation to introduce negligible uncertainty in the theoretical prediction. In this work we adopt the most used definition of the effective redshift
\begin{equation}\label{eq:def_zeff}
    z_{\rm eff} =
    \frac{
	    \displaystyle
	    \int \dd^3 \vs_1  \,
	   \bar z(\vs_1) \,
	     W^2(\vs_1)
    }{
        \displaystyle
        \int \dd^3 \vs_1 \,
        W^2(\vs_1)
    } \,.
\end{equation}

\section{Results}\label{sec:results}
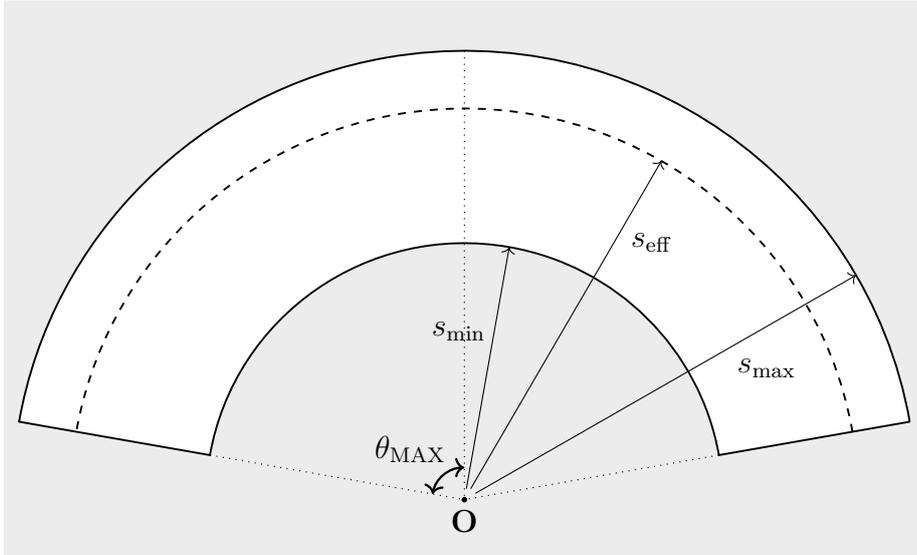
\begin{figure}
\centering
    \begin{tikzpicture}[scale=0.85,
        background rectangle/.style={fill=gray!15},
        show background rectangle
                        ]

        \def\Runo{4}
        \def\Rdue{7}
        \def\Reff{6.1}
        \def\s{0.3*\Reff}
        \def\mispunto{1pt}

        \def\thet{80}
        \def\costheta{ 0.17364817766693041 }
        \def\sintheta{ 0.984807753012208 }

        \def\xpuntouno{\Rdue}
        \def\ypuntouno{0.0}
        \node [name=O] at (\xpuntouno,\ypuntouno) {};

        \def\alphadue{30}
        \def\cosalphadue{ 0.8660254037844387 }
        \def\sinalphadue{ 0.5 }
        \def\xpuntodue{\Rdue + \Rdue * \cosalphadue}
        \def\ypuntodue{\Rdue * \sinalphadue}

        \def\alphatre{80}
        \def\cosalphatre{ 0.17364817766693041 }
        \def\sinalphatre{ 0.984807753012208 }
        \def\xpuntotre{\Rdue + \Runo * \cosalphatre}
        \def\ypuntotre{\Runo * \sinalphatre}

        \def\alphaquattro{60}
        \def\cosalphaquattro{ 0.5 }
        \def\sinalphaquattro{ 0.8660254037844387 }
        \def\xpuntoquattro{\Rdue + \Reff * \cosalphaquattro}
        \def\ypuntoquattro{\Reff * \sinalphaquattro}

        \fill [white] (\xpuntouno, \ypuntouno) -- (\xpuntouno + \sintheta * \Rdue, \costheta * \Rdue) arc (90-\thet:90+\thet:\Rdue) -- (\xpuntouno, \ypuntouno);

        \fill [gray!15] (\xpuntouno, \ypuntouno) -- (\xpuntouno +\sintheta * \Runo, \costheta * \Runo) arc (90-\thet:90+\thet:\Runo) -- (\xpuntouno, \ypuntouno);

        \draw [line width=0.25mm] (\xpuntouno + \sintheta * \Rdue, \costheta * \Rdue)
            arc (90-\thet:90+\thet:\Rdue) --
            (\xpuntouno - \sintheta * \Runo, \costheta * \Runo)  --
            (\xpuntouno - \sintheta * \Runo, \costheta * \Runo) arc (90+\thet:90-\thet:\Runo) --
            (\xpuntouno + \sintheta * \Rdue, \costheta * \Rdue);
        \draw [line width=0.25mm, dashed] (\xpuntouno + \sintheta * \Reff, \costheta * \Reff)
            arc (90-\thet:90+\thet:\Reff);

        \draw[line width = 0.15mm, dotted] (\xpuntouno - \sintheta * \Runo, \costheta * \Runo) -- (\xpuntouno, \ypuntouno);
        \draw[line width = 0.15mm, dotted] (\xpuntouno + \sintheta * \Runo, \costheta * \Runo) -- (\xpuntouno, \ypuntouno);
        \draw[line width = 0.15mm, dotted] (\xpuntouno, \Rdue) -- (\xpuntouno, \ypuntouno);

        \def\radcicletheta{0.5}
        \draw[line width = 0.3 mm, <->] (\xpuntouno - \sintheta * \radcicletheta, \costheta * \radcicletheta) arc (90+\thet:90:\radcicletheta);
        \draw[above] node(te) at  (\xpuntouno - 1.7 * \sintheta * \radcicletheta, 4.7*\costheta * \radcicletheta) {$\theta_\mathrm{MAX}$};

        \draw[below] node at (O) {$\mathbf{O}$};
        \filldraw[black, above] (O)circle (\mispunto);

        \draw[->] (O) to (\xpuntodue, \ypuntodue);
        \draw[->] (O) to (\xpuntotre, \ypuntotre);
        \draw[->] (O) to (\xpuntoquattro, \ypuntoquattro);

        \draw[left] node at (\xpuntouno + 0.5, 0.75*\ypuntodue) {$s_\mathrm{min}$};
        \draw[left] node at (\xpuntouno + 5.3, 0.52*\ypuntotre) {$s_\mathrm{max}$};
        \draw[above] node at (\xpuntouno + 2.9, 0.7*\ypuntoquattro) {$s_\mathrm{eff}$};

    \end{tikzpicture}
    \caption{Transversal cut of the toy-model survey volume considered, which is an azimuthally symmetric spherical cap included between the comoving distances $s_\mathrm{min}$ and $s_\mathrm{max}$ and with angular opening around the central axis $\theta_\mathrm{MAX}$. We set $\theta_{\rm MAX}$ to cover approximately 1/2 of the sky. The dashed line shows the effective comoving distance $s_\mathrm{eff}$.}
    \label{fig:grafico_survey_1}
\end{figure}

In this Section we want to quantify the importance of wide-angle and GR corrections to the measured power spectra of luminosity distance fluctuations. The only missing ingredient is the shape of the window function.
Our code GaPSE~\cite{Foglieni:2023xca} can handle arbitrary complicated window functions in input, however here for simplicity we assume an azimuthally symmetric window, whose multipoles $Q_\ell(s)$ can be computed analytically (see Appendix A of \cite{Casto_DiDio}). The toy window function used in this work is shown in Fig.~\ref{fig:grafico_survey_1}. We assume the data cover 1/2 of the sky and measure a constant number of objects between a lower redshift $z_{\rm min}$ and a higher redshift $z_{\rm max}$. In our baseline scenario $z_{\rm min} = 0.05$ and $z_{\rm max} = 0.2$.

In the plane-parallel limit, linear theory, and neglecting GR corrections, the auto-correlation of luminosity distance fluctuations reads, with $\mu_s = \hat{\mathbf{s}}\cdot \hat{\mathbf{s}}_1$,
\begin{align}
    \xi^{(vv)} (s, \mu_s,z_{\rm eff}) & =
            f^2(z_{\rm eff}) \mathcal{H}^2(z_{\rm eff})
            D^2(z_{\rm eff}) \, s^2
            \left[
                \frac{1}{3} I^2_0(s) \mathcal{L}_0(\mu_s) -
                \frac{2}{3} I^2_2(s) \mathcal{L}_2(\mu_s)
            \right]  \,,
\end{align}
while the cross-correlation between luminosities and galaxies is
\begin{align}
         \xi^{(\delta v)} (s, \mu_s,z_{\rm eff}) =&
            f(z_{\rm eff}) \mathcal{H}(z_{\rm eff})
            D^2(z_{\rm eff}) \, s \; \times
            \notag \\
            &\left[
                - \left(b+\frac{3}{5}f(z_{\rm eff})\right)
                I^1_1(s) \mathcal{L}_1(\mu_s) +
                \frac{2}{5} f(z_{\rm eff})
                I^1_3(s) \mathcal{L}_3(\mu_s)
            \right]\,,
\end{align}
where $b$ is the effective linear galaxy bias of the sample.
In the above expressions, and for consistency with the literature, we have added the superscript $(vv)$ to the auto-correlation of luminosity distances, and  a superscript $(\delta v)$ for their cross-correlation with the galaxy number density. This is motivated by the fact that, in the flat-sky limit and from a simple counting of the derivatives appearing in \eqref{eq:fluctuations_ld}, the correlation functions of luminosity distances are dominated, at small scales, by the radial peculiar velocities, $v_{||}$ appearing in \eqref{eq:fluctuations_ld}. In the remainder of this work we will use the terminology `luminosity distances' and `peculiar velocities' interchangeably when no confusion can possibly arise.
We see that the auto-correlation function contains a monopole and a quadrupole term, while the cross-correlation function is expanded into a dipole and a octupole piece. In Fourier space, under the same assumption as above, and also neglecting the effect of the window function, one has
\begin{align}
\label{eq:P_vv}
    P^{(vv)} (k, \mu_k,z_{\rm eff})  =
            f^2(z_{\rm eff}) \frac{\mathcal{H}^2(z_{\rm eff})}{k^2}
            D^2(z_{\rm eff}) P(k)\,
            \left[
                \frac{1}{3} \mathcal{L}_0(\mu_k) -
                \frac{2}{3} \mathcal{L}_2(\mu_k)
            \right]  \,,
\end{align}
and
\begin{align}
\label{eq:P_dv}
         P^{(\delta v)} (k, \mu_k,z_{\rm eff}) =&
            i f(z_{\rm eff}) \frac{\mathcal{H}(z_{\rm eff})}{k}
            D^2(z_{\rm eff}) P(k)\,  \; \times
            \notag \\
            &\left[
                - \left(b+\frac{3}{5}f(z_{\rm eff})\right)
                 \mathcal{L}_1(\mu_k) +
                \frac{2}{5} f(z_{\rm eff})
                 \mathcal{L}_3(\mu_k)
            \right]\,,
\end{align}
where $\mu_k = \vb{\hat{k}}\cdot \vb{\hat{s}_1}$, is the cosine of the angle between the wavenumber $ \vb{\hat{k}}$ and the chosen line of sight direction $\vb{\hat{s}_1}$. This is the so called local-parallel approximation\footnote{Replacing $\hat{\mathbf{s}}_1$ with some fixed direction $\vb{\hat{z}}$ leads to the global plane-parallel approximation, leaving the analytical expressions~(\ref{eq:P_vv},~\ref{eq:P_dv}) unchanged.}.
Notice the cross-correlation power spectrum,~\eqref{eq:P_dv}, is purely imaginary.

Wide angle and GR terms will manifest as correction to the above expressions and as new terms in the multipole expansion. Specifically, for wide-angle contributions we indicate any correction to the above equations, \emph{i.e.}~beyond the flat-sky limit, coming from the auto-correlation of radial velocities in the case of $P^{(vv)}$, and from the cross correlation between densities and velocities for $P^{(\delta v)}$. Everything else beyond wide-angle terms is then dubbed as a GR correction.

\begin{figure}
    \centering
\includegraphics[width=0.8\textwidth]{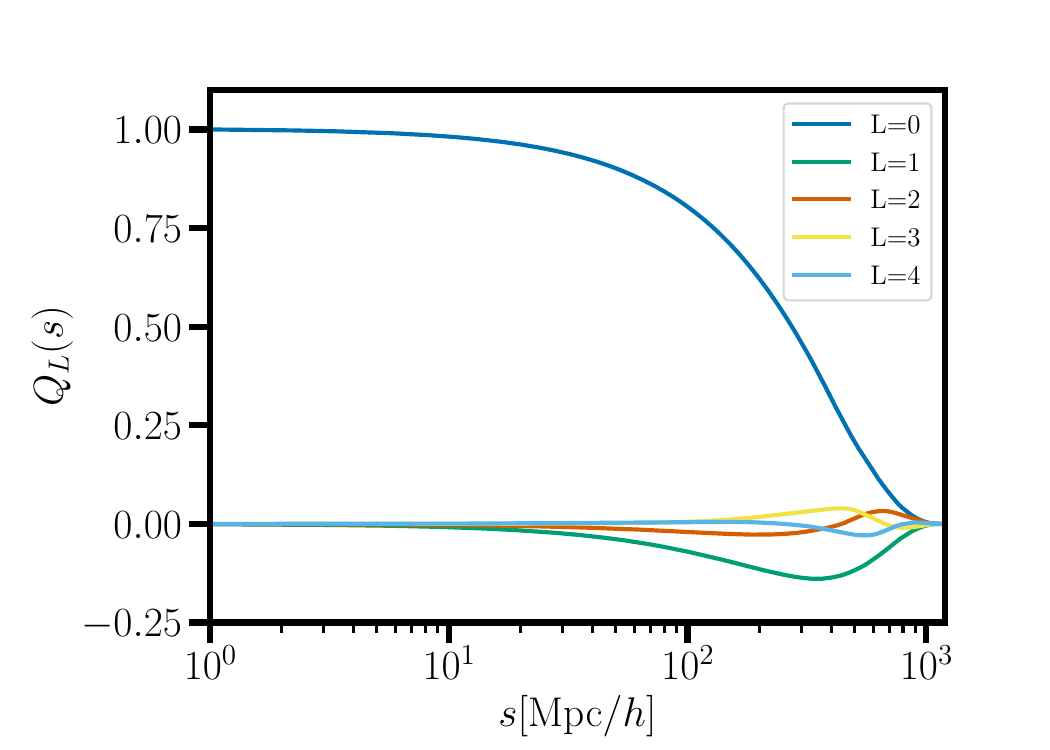}
    \caption{The multipoles of our toy-model window function up to $L=4$, normalized to $Q_0(s\rightarrow 0) = 1$. The window function, as shown in Fig.~\ref{fig:grafico_survey_1}, is defined as $W(\mathbf{s}) = \Theta(s - s_\mathrm{min} )\Theta(s_\mathrm{max} - s) \Theta(\theta_\mathrm{max} - \theta)$ where we take $s_\mathrm{min} = s(z_\mathrm{min}=0.05)$,  $s_\mathrm{max} = s(z_\mathrm{max}=0.2)$ and $\theta_\mathrm{max} = \pi/2$.}
    \label{fig:window_multipoles}
\end{figure}

\begin{figure}
    \centering
\includegraphics[width=0.32\textwidth]{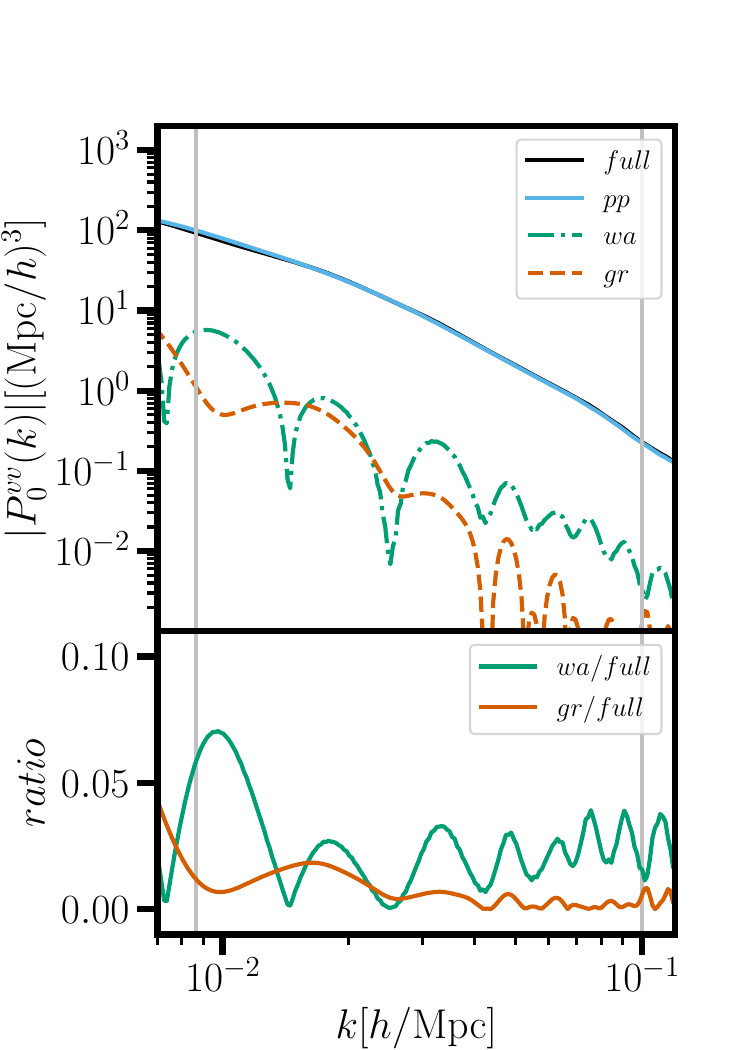}
\includegraphics[width=0.32\textwidth]{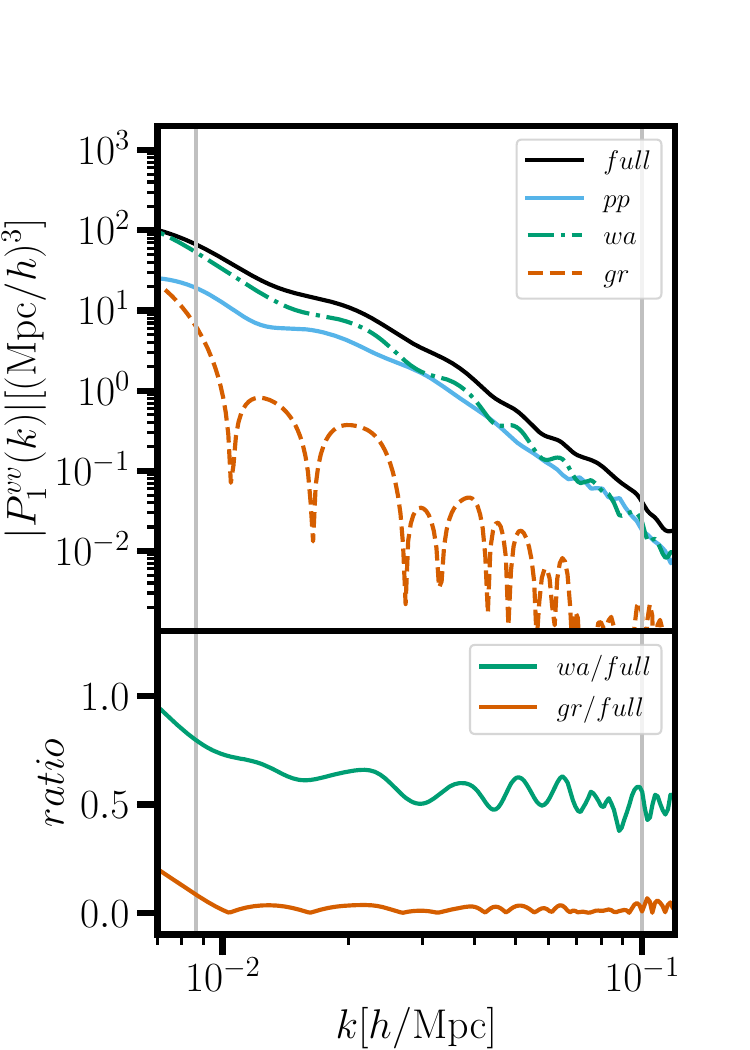}
\includegraphics[width=0.32\textwidth]{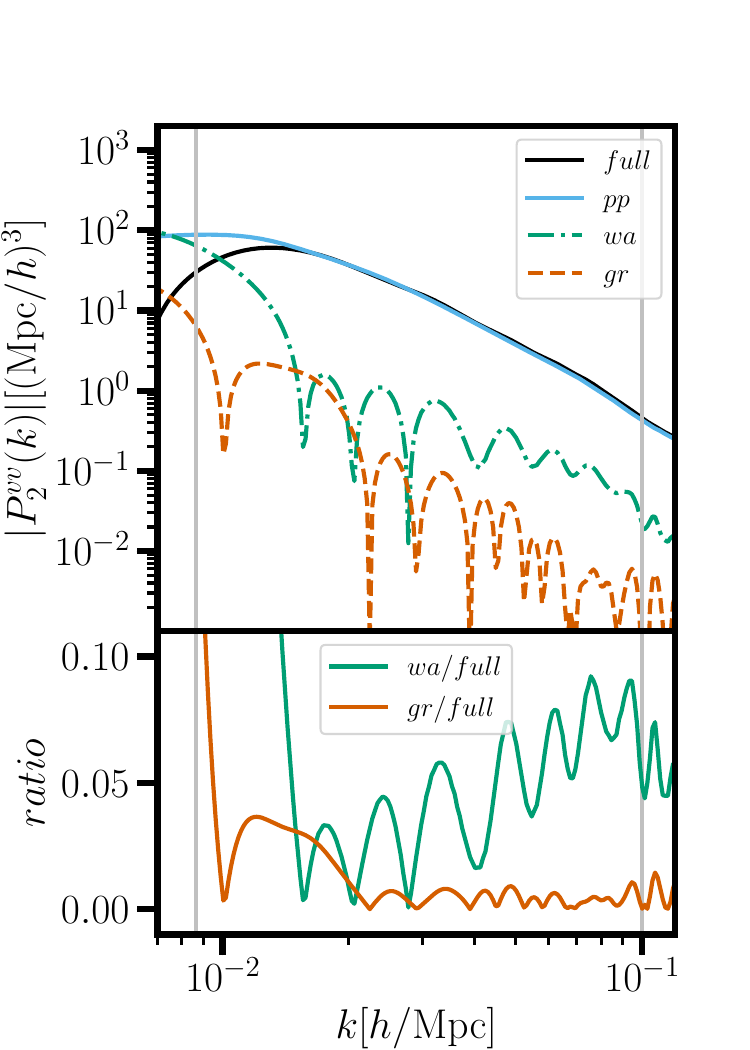}
\includegraphics[width=0.32\textwidth]{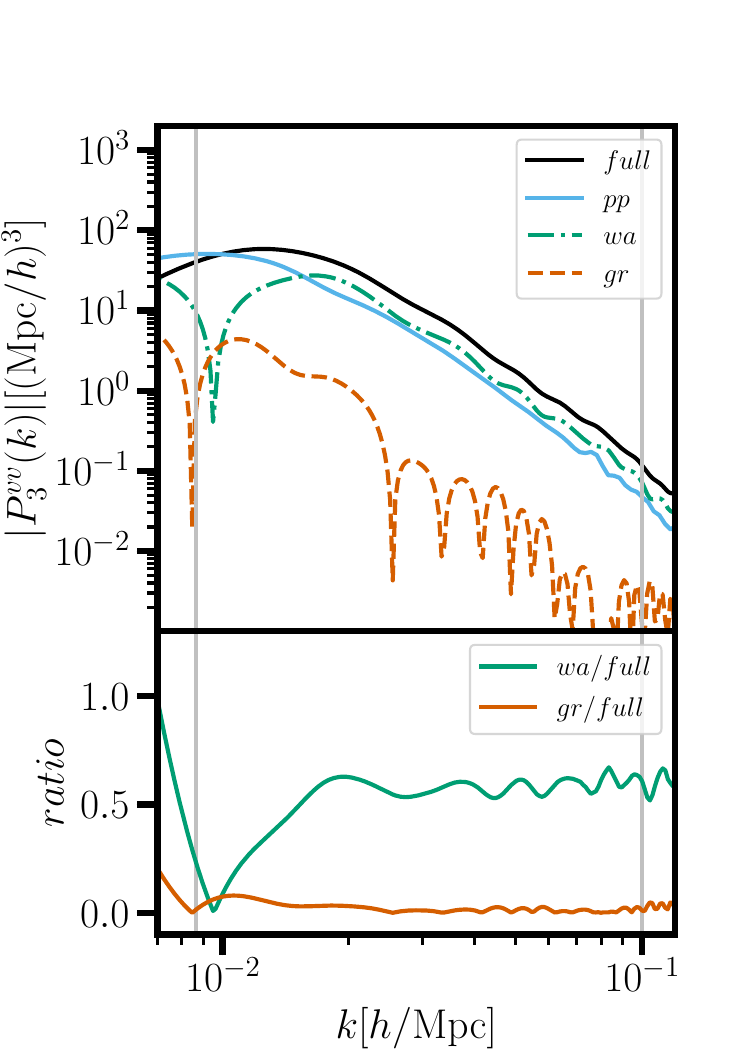}
\includegraphics[width=0.32\textwidth]{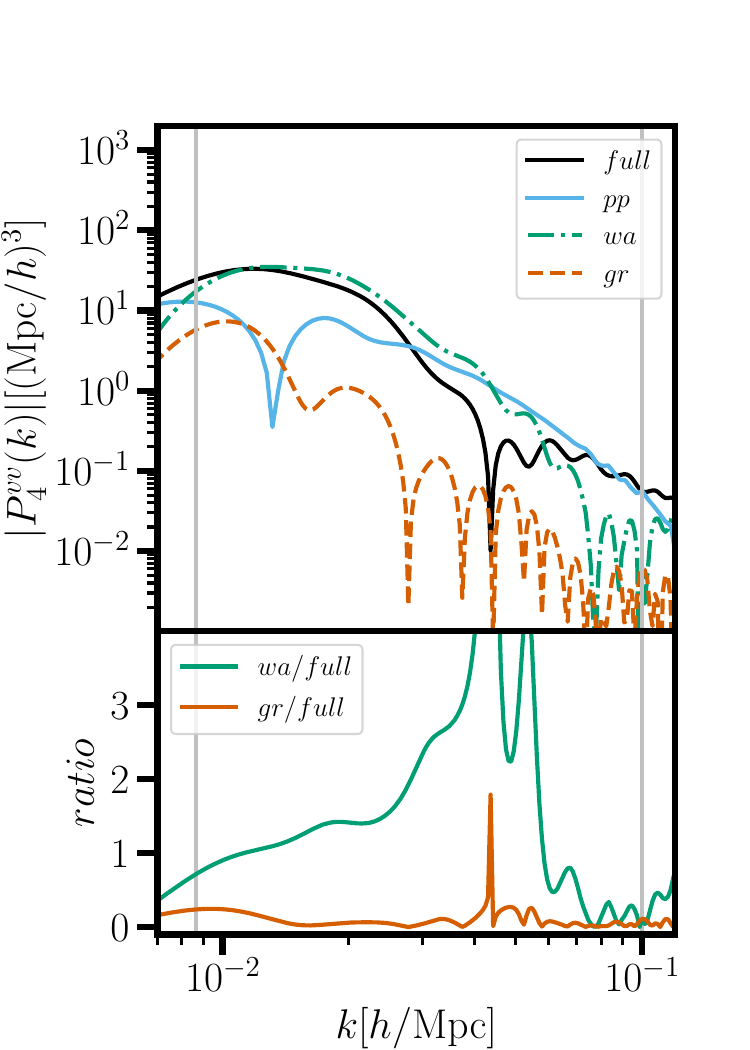}
    \caption{The multipoles of the luminosity distance power spectrum $P^{(vv)}_L(k)$ up to $L=4$. We consider a survey covering half of the sky between $z_{\rm min} = 0.05$ $z_{\rm max} = 0.2$ with a constant number density. In all panels the full theory expressions is shown with continuous black lines, the plane-parallel limit in cyan, the wide-angle term with dot-dashed green lines, and the GR term in orange dashed lines. In each plot, the lower panels displays the ratio between wide-angle or GR terms and the full expression for the multipoles.}
    \label{fig:Pvv}
\end{figure}

Fig.~\ref{fig:Pvv} shows the first five multipoles of the luminosity distance fluctuations power spectrum for the baseline survey configurations discussed at the beginning of this section and plotted in Fig.~\ref{fig:grafico_survey_1}. The two vertical lines indicate $k = 0.1 \,h/$Mpc and the largest scale available in the survey $k_{\rm min} \sim 2\pi/V^{1/3} \sim 0.009 \,h/$Mpc, where $V$ is the survey volume.

For the power spectrum monopole $P_0^{(vv)}(k)$, shown in the top-left panel, we see that the plane-parallel term, in cyan, dominates over the others on all scales. Wide angle corrections are shown in green, while GR corrections in orange, and contribute at most a few \% of the signal at edge of the survey. The other non-zero multipole in the plane parallel limit is $P_2^{(vv)}$, shown
on the top right in Fig.~\ref{fig:Pvv}. The flat-sky term is indeed dominant on most scales, but towards the survey boundary, and for our choice of survey window function, the wide angle corrections, in green, can become comparable to the flat-sky contribution. The GR corrections too can reach up to 10\% of the overall amplitude of the quadrupole of peculiar velocities at low-$k$, and they are dominated by the cross correlations between peculiar velocities and the local gravitational potential.

The dipole, octopule and hexadecapole would instead be zero in the flat-sky linear theory and in the absence of a window function. The dipole $P_{1}^{(vv)}(k)$ is shown in the top center panel in Fig.~\ref{fig:Pvv}. The contribution in the plane-parallel limit is coming from the coupling induced by the window function (whose multipoles are shown in Fig.~\ref{fig:window_multipoles}), which causes part of the power in the monopole and quadrupole to leak into other multipoles. The size of this term is indeed one order of magnitude smaller than the one of the first even multipoles. The wide-angle piece is comparable in amplitude to the plane-parallel term on all scales and it dominates the total dipole at large scales. The GR corrections, mostly sourced by the cross correlations between velocities and the local gravitational potential, are small but can reach about 10\% of the total signal at low-$k$.
Similar conclusions apply to the octopule and the hexadecapole.

\begin{figure}
    \centering
\includegraphics[width=0.32\textwidth]{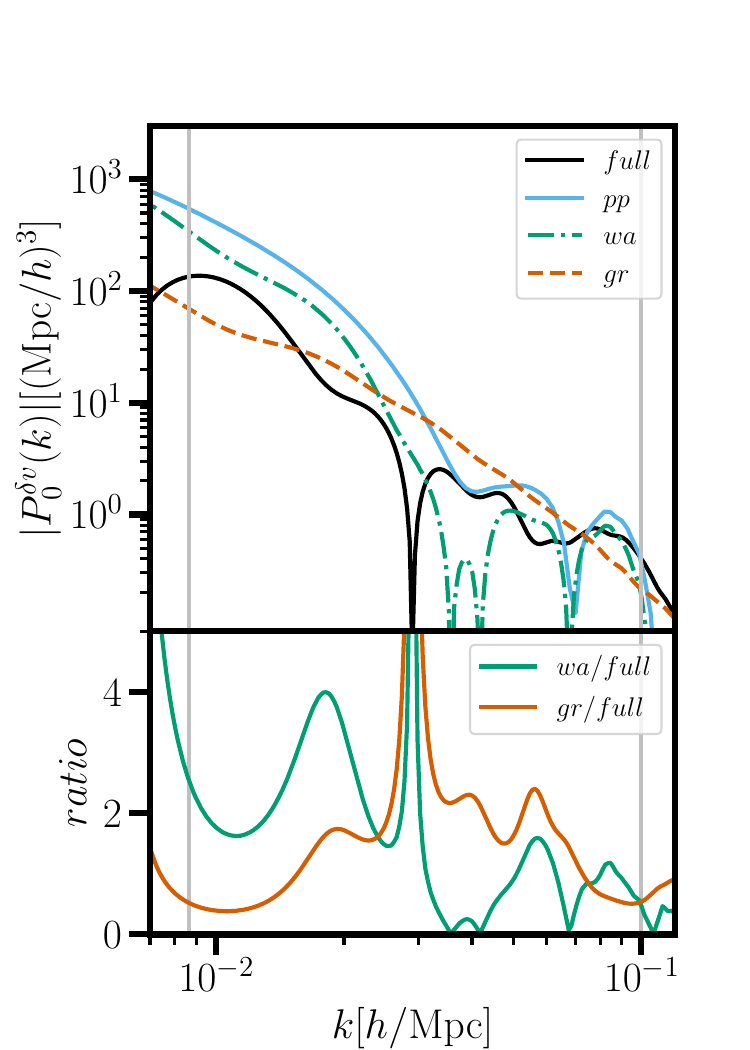}
\includegraphics[width=0.32\textwidth]{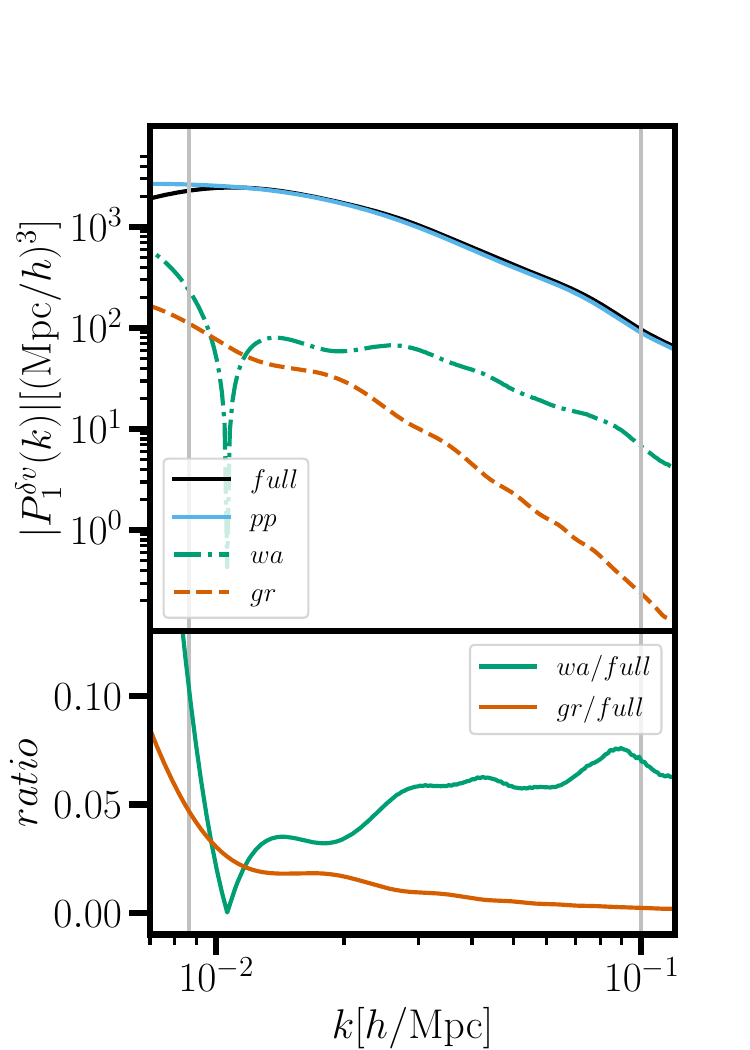}
\includegraphics[width=0.32\textwidth]{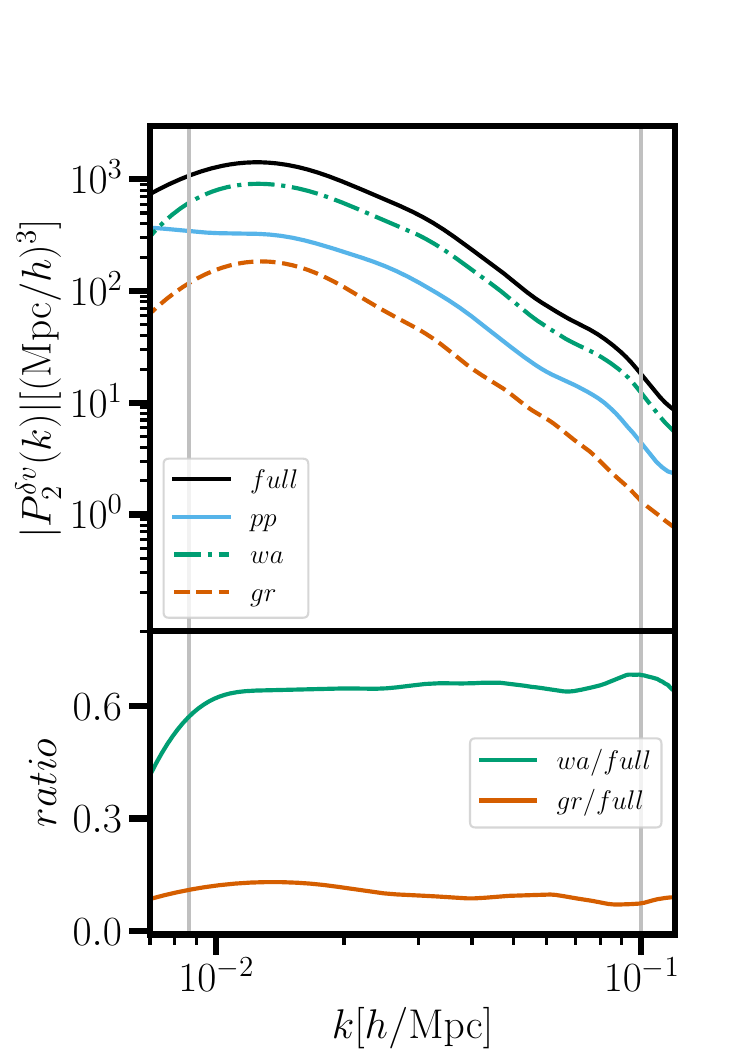}
\includegraphics[width=0.32\textwidth]{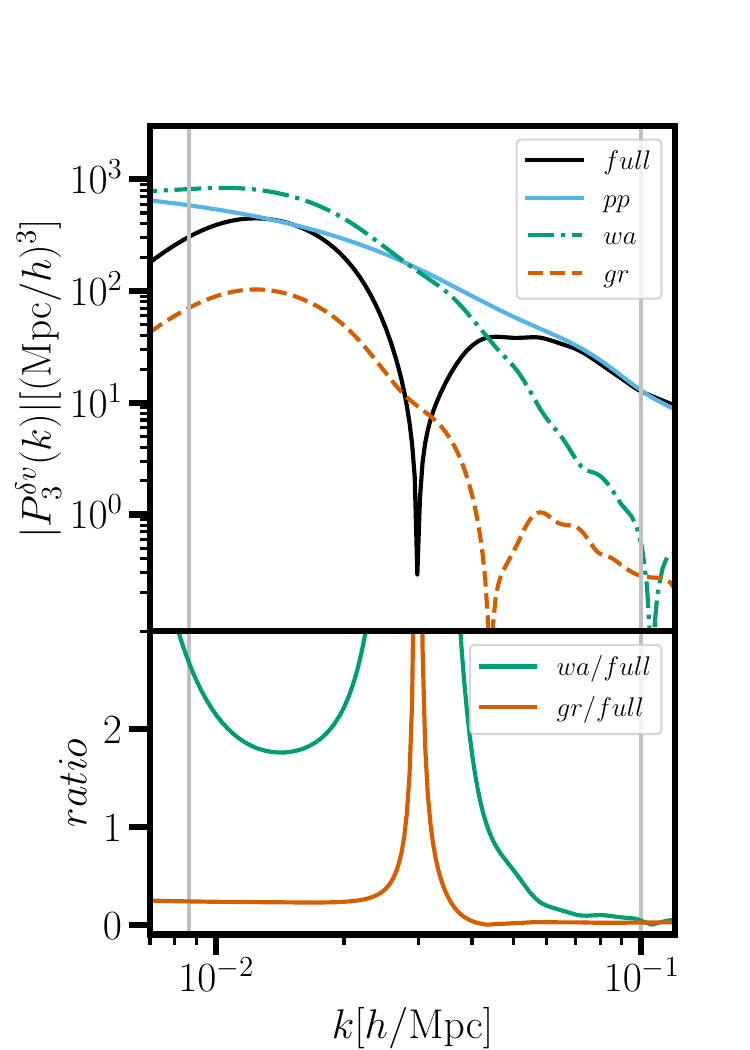}
\includegraphics[width=0.32\textwidth]{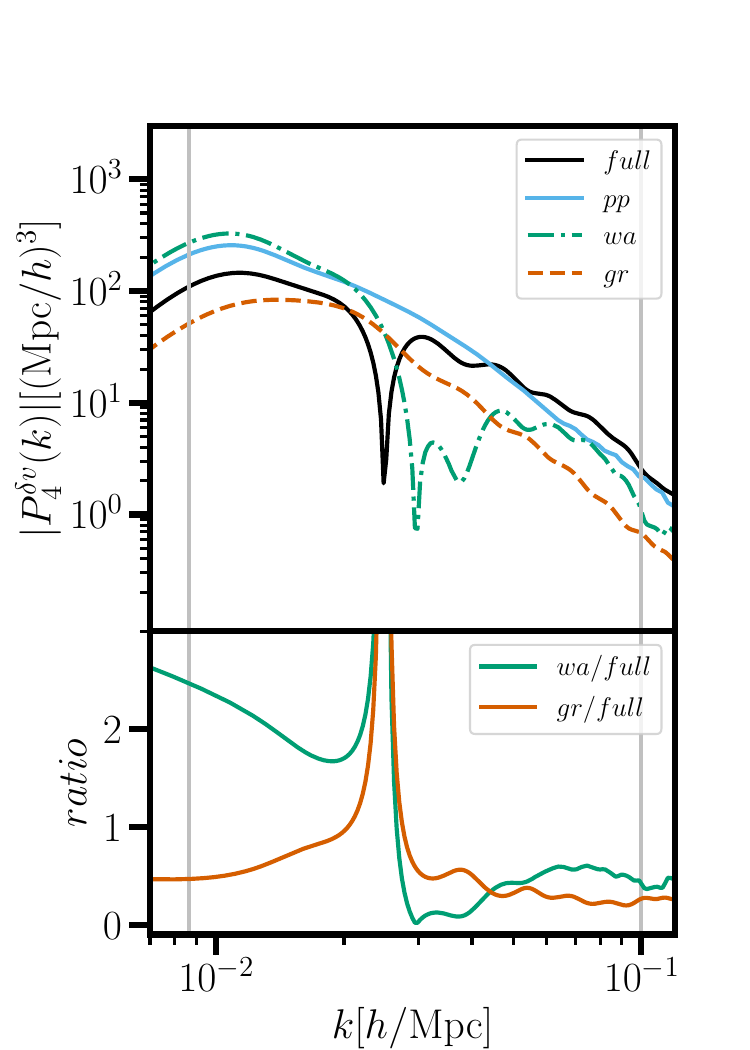}
    \caption{The multipoles of the cross-power spectrum between luminosity distance fluctuations and galaxy number counts $P^{(\delta v)}_L(k)$ up to $L=4$. We consider a survey covering half of the sky between $z_{\rm min} = 0.05$ $z_{\rm max} = 0.2$ with a constant number density. In all panels the full theory expressions is shown with continuous black lines, the plane-parallel limit in cyan, the wide-angle term with dot-dashed green lines, and the GR term in orange dashed lines. In each plot, the lower panels displays the ratio between wide-angle or GR terms and the full expression for the multipoles.
    }
    \label{fig:Pdv}
\end{figure}

We now turn our attention to the multipoles of the cross-power spectrum between luminosity distances and galaxy number counts, shown in in Fig.~\ref{fig:Pdv}. In this case, the leading term in the plane-parallel limit is the dipole $P_1^{(\delta v)}$, shown in the center panel of the upper row. As expected, the flat-sky expression in \eqref{eq:P_dv}, shown as the cyan line, is the largest contribution to the total dipole. Wide-angle corrections are small but not completely negligible, being roughly 5\% at small scales and up to 10\% at the survey boundary. GR terms instead are below 5\% on all scales and can therefore be safely neglected.
The octopule $P^{(\delta v)}_3(k)$ is shown in the bottom left panel in Fig.~\ref{fig:Pdv}. The flat-sky limit is the leading term at $k\gtrsim 0.05\,h/$Mpc, but it receives large wide-angle corrections at small wavenumbers. The two have actually opposite signs, leading to a significant cancellation of the total octopule. This in turn makes the GR corrections more important, as they can reach 20\% of the signal at low-$k$.

Even multipoles in the cross-correlations are instead zero in the plane-parallel limit and neglecting the presence of a window function. This is only an approximation, for instance the latter is responsible for the cyan set of curves in Fig~\ref{fig:Pdv} for the multipoles with $\ell=0,\,2,\,4$. Considering the monopole $P_0^{(\delta v)}(k)$, shown in the top-left panel, the wide-angle and flat-sky terms have opposite sign and a similar amplitude, resulting in a large cancellation. The GR correction, the dashed orange curve, is actually $\mathcal{O}(1)$ of the total signal all the way from the survey boundary to $k \sim 0.1\,h/$Mpc. The quadrupole $P^{(\delta v)}_2(k)$, shown in the top-right panel of Fig.~\ref{fig:Pdv}, is primarily sourced by wide-angle and GR terms, accounting roughly for the 60\% and 10\% of the total quadrupole respectively.

Finally, for the hexadecapole $P^{(\delta v)}_4(k)$, shown in the bottom-right panel, we once again see the partial cancellation between the flat-sky and wide-angle contributions on large scales,  with GR terms being a substantial $\mathcal{O}(1)$ correction on all scales.

\subsection{Detectability of wide-angle and GR effects}\label{sec:forecasts}
Equipped with the prediction of the multipoles of the luminosity distance power spectrum and their cross-correlation with the galaxy density field, we can now forecast the detection significance of wide-angle and GR corrections.

We achieve this with a Fisher matrix approach. Given a set of observables $\boldsymbol{\mu}$, \emph{e.g.} power spectrum multipoles, and their covariance matrix $\mathbf{C}$, the Fisher matrix between two parameters $\theta_i$ and $\theta_j$ is defined as

\begin{align}
    F_{ij} = \frac{V}{4 \pi^2} \int_{k_{\rm min}}^{k_{\rm max}} \dd k\,k^2 \frac{\partial
    {\boldsymbol{\mu}}}{\partial \theta_i}\mathbf{C}^{-1}\frac{\partial
    {\boldsymbol{\mu}}}{\partial \theta_j}\,,
\end{align}
where $V$ is the survey volume, and $k_{\rm min}$ and $k_{\rm max}$ are the smallest and largest scale included in the data analysis respectively. The marginalized uncertainty on a given parameter is then defined as $\sigma_{\theta_i} = \sqrt{(F^{-1})_{ii}}$. The data vector we consider is
\begin{align}
\label{eq:mu}
    \boldsymbol{\mu} = \{P_0^{(\delta \delta)}+\frac{1}{\bar{n}_g},\,P_2^{(\delta \delta)},\,P_4^{(\delta \delta)},\,P_0^{(\delta v)},\,P_1^{(\delta v)},P_2^{(\delta v)}\,\,P_3^{(\delta \delta)},P_4^{(\delta v)},\,P_0^{(v v)} + \frac{\sigma^2}{\bar{n}_v},\,P_2^{(vv)},\,P_4^{(vv)}\}\,,
\end{align}
where the $P_{L}^{(\delta \delta)}(k)$ are the multipoles of the galaxy power spectrum, also predicted using GaPSE \cite{Foglieni:2023xca}. We included a shot-noise contribution to the galaxy power spectrum monopole going as $1/\bar{n}_g$, where $\bar{n}_g$ is the galaxies mean number density. The shot-noise term in the monopole of the velocity power spectra depends on the number density of galaxies for which a measurement of the luminosity distance is available, $\bar{n}_v$, and on the scatter of each individual measurement $\sigma^2$. Following previous work in the literature, \emph{e.g.} \cite{whitford2022using}, we model this dispersion as
\begin{align}
\label{eq:alpha}
    \sigma^2 = \sigma^2_{\rm rand} + [\alpha H_0 \chi(z)]^2\,,
\end{align}
where $\sigma^2_{\rm rand} = 300$ km/s accounts for the non-linear contribution to the peculiar velocities, and the second terms reflects the intrinsic scatter in the reconstructed luminosity distances. For the FP or the TF relation, the fractional uncertainty in the measured distance of each object roughly corresponds to $\alpha \simeq 0.2$ \cite{howlett2017cosmological}, while it decreases to $\alpha \simeq 0.1$ for SN1A distance measurements.

The careful reader might have noted that our data vector in \eqref{eq:mu} does not include the odd multipoles of the auto-power spectra. This is due to the fact that while the expressions for the signal beyond the plane-parallel limit are available and implemented in GaPSE, the same cannot be said for the covariance between power spectrum multipoles beyond the flat-sky limit. Given four tracers $X$, $Y$, $W$ and $Z$, the flat-sky covariance between the cross-power spectrum of $X$ and $Y$, $\hat P^{(XY)}(\mathbf{k}_1)$, and the cross-power spectrum of $W$ and $Z$, $\hat P^{(WZ)}(\mathbf{k}_2)$, reads
\begin{align}
   &  \text{Cov}[\hat P^{(XY)}(\mathbf{k}_1) \hat P^{(WZ)}(\mathbf{k}_2)]  \notag\\
    & \equiv \langle \delta_X(\mathbf{k}_1) \delta_Y(\mathbf{-k}_1) \delta_W(\mathbf{-k}_2) \delta_Z(\mathbf{k}_2) \rangle - \avg{\delta_X(\mathbf{k}_1) \delta_Y(\mathbf{-k}_1)}\avg{\delta_W(\mathbf{-k}_2) \delta_Z(\mathbf{k}_2)} \notag\\
    & = \langle \delta_X(\mathbf{k}_1) \delta_W(\mathbf{-k}_2) \rangle \langle \delta_Y(\mathbf{-k}_1) \delta_Z(\mathbf{k}_2) \rangle + \langle \delta_X(\mathbf{k}_1) \delta_Z(\mathbf{k}_2) \rangle \langle \delta_Y(\mathbf{-k}_1) \delta_W(\mathbf{-k}_2) \rangle  \notag\\
    & =  P^{(XW)}(\mathbf{k}_1)\delta_D^{(3)}(\mathbf{k}_1 - \mathbf{k}_2)P^{(YZ)}(\mathbf{-k}_1)\delta_D^{(3)}(\mathbf{k}_1 - \mathbf{k}_2)  \notag\\   & +  P^{(XZ)}(\mathbf{k}_1)\delta_D^{(3)}(\mathbf{k}_1 + \mathbf{k}_2)P^{(YW)}(\mathbf{-k}_1)\delta_D^{(3)}(\mathbf{k}_1 + \mathbf{k}_2) \notag \\
    & \equiv   \mathcal{C}^{(XYWZ)}(\mathbf{k}_1, \mathbf{k}_2)\,.
    \label{eq:covariance_definition}
\end{align}
In our case the four tracers are either the galaxy number counts or the fluctuations in the luminosity distance. Actually, we are interested in the covariance between the multipoles of the power spectra, $P_{L_1}^{(XY)}(k_1)$ and $P_{L_2}^{(WZ)}(k_2)$, which amounts to a simple projection of the equation above. The final result is, see App.~\ref{app:Covariance_derivation} for a derivation,
\begin{equation}
\label{eq:cov}
    \begin{split}
        \mathcal{C}_{L_1L_2}^{(XYWZ)} (k_1, k_2) & =
        \frac{2L_1+1}{2}(2L_2+1) \frac{\delta_D(k_1-k_2)}{k_1^2} \delta_D^{(3)} (0)
        \\
        &
        \sum_{\ell_1, \ell_2, n_1} (-1)^{\ell_2}\bigg(
        \begin{matrix}
        \ell_1 & L_1 & n_1 \\
        0 & 0 & 0
        \end{matrix}
        \bigg)^2\bigg(
        \begin{matrix}
        \ell_2 & L_2 & n_1 \\
        0 & 0 & 0
        \end{matrix}
        \bigg)^2   \\
        & \times (2n_1+1)[P_{\ell_1}^{(XW)}(k_1) P_{\ell_2}^{(YZ)}(k_1) + (-1)^{L_2} P_{\ell_1}^{(XZ)}(k_1) P_{\ell_2}^{(YW)}(k_1)]
        \\
        &= \frac{\delta_D(k_1-k_2)}{k_1^2} \delta_D^{(3)} (0) {C}^{(XYWZ)}_{L_1 L_2}
        \, .
    \end{split}
\end{equation}
It is easy to see that that flat-sky covariance vanishes for the odd multipoles of the auto-power spectrum of galaxy number counts and of luminosity distance fluctuations, \emph{i.e.} for a single tracer.
This is a technical limitation, not only of this work, but in general of the modeling of observables beyond the plane-parallel limit, and we intend to return to this issue in future work \footnote{Recent work in \cite{Wen:2024hqj} presents a possible solution, although with a large computational cost.}. In this respect our Fisher matrix estimates are conservative, as we do not include all possible signals in our analysis.

The parameters we are interested in are the cumulative size of the wide-angle and GR corrections to a set of measurements of the galaxy and velocity power spectra. These are defined by taking each multipole $P_\ell^{(XY)}(k)$ and splitting it into a plane-parallel, a wide-angle, and a GR term as follows,
\begin{align}
    P_\ell^{(XY)}(k) = P_{\ell, pp}^{(XY)}(k) + A P_{\ell, wa}^{(XY)}(k)+BP_{\ell, GR}^{(XY)}(k)\,,
\end{align}
with a fiducial value of 1 for both $A$ and $B$. The three terms defined above correspond to the cyan, green and orange curves we have shown in Figs.~\ref{fig:Pvv} and \ref{fig:Pdv}. We then forecast the errorbars of $A$ and $B$ around their fiducial value and for a given survey configuration. A marginalized uncertainty of $\sigma_A = 0.2$ thus means wide-angle effects will be detected at 5-$\sigma$.
The procedure outlined above is equivalent to computing the total signal-to-noise in wide-angle and GR corrections to the measured multipoles.
\begin{figure}
     \centering
     \includegraphics[width=0.495\textwidth]{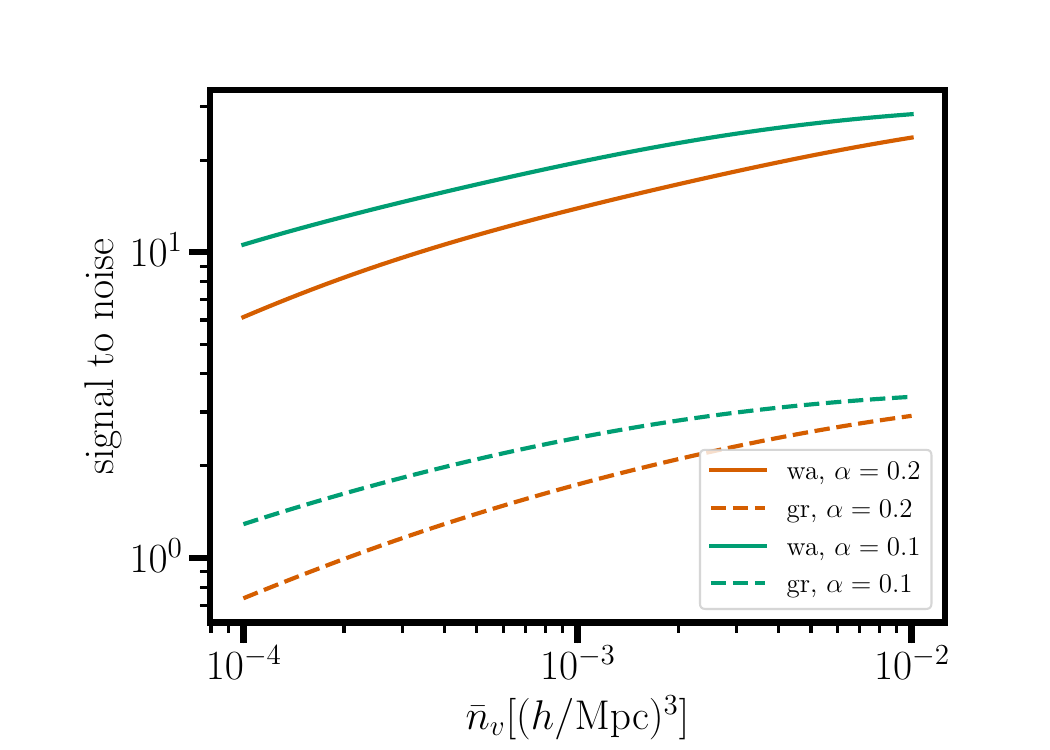}
     \includegraphics[width=0.495\textwidth]{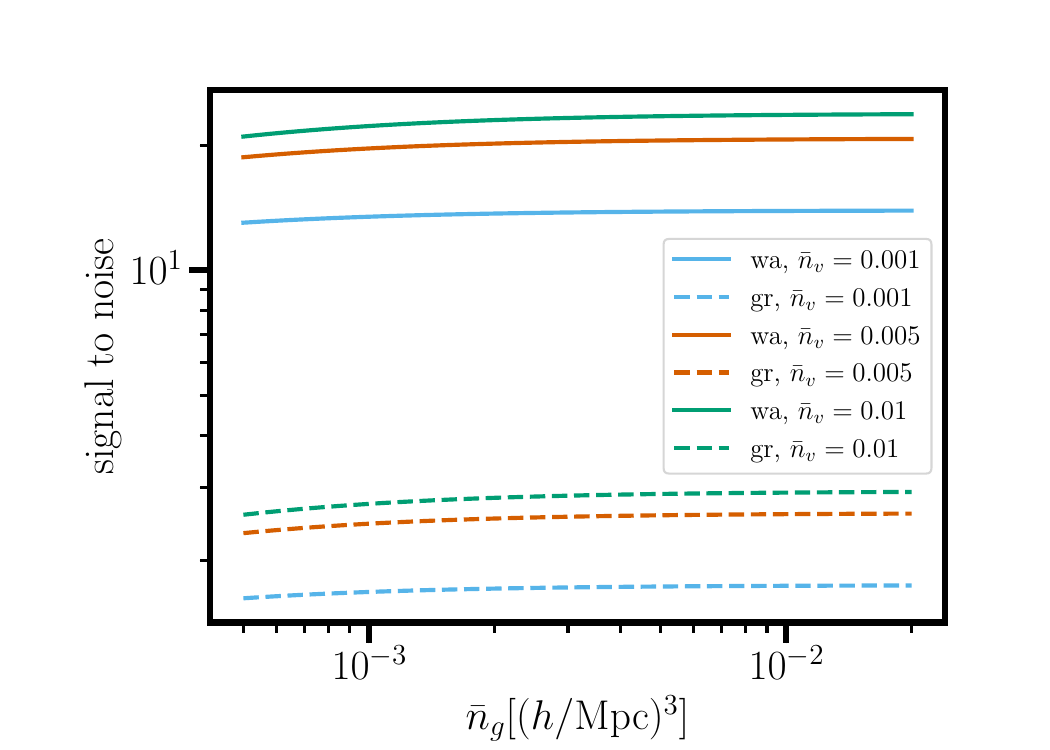}
     \caption{Comulative signal-to-noise in wide-angle and GR corrections. In all cases, fiducial parameters are fixed to the values in Eq.~\ref{eq:fid}, and we included all multipoles in Eq.~\ref{eq:mu}. \emph{Left panel:} Signal to noise in wide-angle terms (continuous lines) and GR terms (dashed lines) as a function of the number of galaxy with an associated distance measurement, $\bar{n}_v$. Green lines corresponds a scatter in the measurements of distance of $\alpha =0.1$, while the orange line to $\alpha =0.2$, see Eq.~\ref{eq:alpha}. \emph{Right panel:} Total signal to noise as we vary the number density of galaxies $\bar{n}_g$. Different colors corresponds to different values of $\bar{n}_v$, while different dashing to wide-angle or GR effects.}
     \label{fig:Fisher}
 \end{figure}
The fiducial cosmological parameters are set to the Planck 2018 best-fit values \cite{Planck_2018}, while we take the following fiducial values for the survey parameters
\begin{align}
\label{eq:fid}
    & z_{\rm eff} = 0.15\;,\; V = 3.8\times10^8 (\mathrm{Mpc}/h)^3\;,\; b_1 = 1.5\;,\bar{n}_g = 10^{-2} \,(h/\mathrm{Mpc})^3  \\ & \bar{n}_v = 10^{-3} \,(h/\mathrm{Mpc})^3\;,
     \alpha =0.2\;,\; k_{\rm min} =0.026\,h/\mathrm{Mpc} \;,\; k_{\rm max} = 0.1\, h/\mathrm{Mpc} \notag
\end{align}
We will also show results varying some of them. The fiducial values of $\bar{n}_g$  and $\bar{n}_v$ correspond to an optimistic DESI peculiar velocity survey \cite{DESIPV}.
The largest scale included in our Fisher matrix,
$k_{\rm min} = 0.026\,h/$Mpc, does not correspond to the smallest wavenumber available in the survey $\sim 2\pi/V^{1/3}$, since our expression for the covariance of the multipoles in Eq.~\ref{eq:cov} does not include the effect of a window function, which we expect to be more significant at low-$k$. Also in this respect our analysis is conservative, since both wide-angle and GR correction peak at large scales. In all cases we set $k_{\rm max} = 0.1\,h/$Mpc.
Finally, our Fisher matrix is effectively one-dimensional, as we vary either $A$ or $B$ while keeping all the other parameters fixed.

The result of the Fisher matrix calculations are summarized in Fig.~\ref{fig:Fisher}. In the left panel we show the forecasted signal-to-noise of the wide-angle terms, with continuous lines, and of the GR terms, with dashed lines, as we vary the number density of the galaxies for which a measurement of the luminosity distance is available. The orange lines correspond to a scatter in the measurement of the luminosity distance per object of $\alpha = 0.2$, while the green lines correspond to the more optimistic case of $\alpha = 0.1$.
We find that wide-angle effects should be detected at high-significance, with a signal to noise $\gtrsim 10$, regardless of the value of $\bar{n}_v$ and $\alpha$. This was expected given the results we have seen for the multipoles in Figs.~\ref{fig:Pvv} and \ref{fig:Pdv}.

Detecting GR effects is instead more challenging. For $\alpha = 0.2$, a 5-$\sigma$ threshold requires $\bar{n}_v \gtrsim 10^{-2}\,h^3/\text{Mpc}^3$, which is beyond current capabilities.
For $\alpha \sim 0.1$, the typical signal-to-noise exceeds unity for $\bar{n}_v \gtrsim 10^{-4}\,h^3/\text{Mpc}^3$, well within what is expected in future velocity surveys.

The right panel of Fig.~\ref{fig:Fisher} shows the dependence of the detection significance on the number density of the galaxy sample $\bar{n}_g$. For both wide-angle and GR terms the total signal-to-noise does not depend really depend on $\bar{n}_g$, indicating that the multipoles of the galaxy power spectrum are less affected by corrections beyond the flat-sky limit, at least until we restrict ourselves to even multipoles only in the auto-spectra.

\section{Conclusions}\label{sec:conclude}
In this work we have presented a comprehensive study of the large-scale luminosity distance power spectrum and its correlation with galaxy number counts. For the first time, we have computed all relativistic and wide-angle corrections to the multipoles of the luminosity distance power spectra, including observational effects like the presence of a window function. All numerical results have been obtained using the our publicly available GaPSE code \footnote{\href{https://github.com/foglienimatteo/GaPSE.jl}{https://github.com/foglienimatteo/GaPSE.jl}}.

We find that for low-redshift surveys, aiming to measure distances via the Tully-Fisher relation, the Fundamental Plane or type Ia Supernovae, wide-angle contributions are important at several scales and cannot be neglected in future data analyses. Relativistic corrections are instead localized at low-$k$, where they can be as large as wide-angle terms. Detecting GR effects at high significance remains challenging due to the small volume probed by the low-redshift surveys considered in this work.
We find that wide-angle and relativistic effects source odd multipoles in the luminosity distance auto-power spectrum and even ones in the cross-correlations between galaxy number counts and luminosity distances. This could increase the detection significance of both terms, provided an analytical covariance matrix beyond the flat-sky limit becomes available, a topic we intend to return in future work.

While in this work we have focused on Fourier space statistics like the power spectrum, our findings should qualitatively apply to any other two-point statistics of luminosity distances. Of particular interest would be to study how wide-angle and GR effects enter in likelihood approaches to luminosity distances \cite{Adams:2020dzw}. Finally, at low-$z$ non-linear corrections to the matter fields cannot be neglected, and it would be interesting to see their interplay with wide-angle and GR corrections, as recently discussed in \cite{Dam:2023cem}.

\acknowledgments
MP acknowledges support from the NWO and the Dutch Ministry of Education, Culture and Science (OCW) (through ENW-XL Grant OCENW.XL21.XL21.025 DUSC). MF acknowledges the Computational X Support (CXS) group at the Leibniz Supercomputing Centre (LRZ) for its support in the latest developments of GaPSE.jl.
ED~acknowledges funding from the European Research Council (ERC) under the European Union’s Horizon 2020 research and innovation program (Grant agreement No. 863929; project title “Testing the law of gravity with novel large-scale structure observables”)

\appendix

\section{Analytical expression of the General Relativistic TPCFs}
\label{app:GR_TPCFs}

In the following sections we present the Two-Point Correlation Functions (TPCFs) that arise from the General Relativistic (GR) perturbation of the Luminosity Distance (LD), and their cross-correlation with the Galaxy Number Counts (GNC).
We present them in terms of the coordinates system $(s_1, s_2, y = \cos\theta = \versor{s}_1\cdot \versor{s}_2)$.

We remind that $\HH = a H$ is the comoving Hubble parameter. We use the following definitions (some of them were already showed in the main text):
\begin{eqnarray}
    \Delta\chi_1 &=& \sqrt{\chi_1^2 + s_2^2-2\,\chi_1\,s_2\,y}
\\
 \Delta\chi_2 &=& \sqrt{s_1^2 + \chi_2^2-2\,s_1\,\chi_2\,y}
\\
    \Delta\chi &=& \sqrt{\chi_1^2 + \chi_2^2-2\,\chi_1\,\chi_2\,y}
\\
    s&=&\sqrt{s_1^2 + s_2^2 - 2 \, s_1 \, s_2 \, y}
\\
    \sigma_i &=& \int \frac{\dd q}{2\pi^2} q^{2-i} \, P(q)
\\
    I^n_l(s) &=& \int \frac{\dd q}{2\pi^2} q^2\, P(q) \frac{j_l(qs)}{(qs)^n}
\\
    \Rld &=& 1 - \frac{1}{\HH r}   \\
    \Rgnc &=& 5 \mb + \frac{2-5\mb}{\HH r} + \frac{\HHd}{\HH^2} +\fevo
\end{eqnarray}

Whenever a quantity is evaluated in $s_1$ or $s_2$, we remove $(s_{1/2})$ and place a subscript ${}_{1/2}$, for instance:
\begin{align*}
    &D_1 = D(s_1) \; , \quad D_2 = D(s_2) \; , \quad
    f_1 = f(s_1) \; , \quad f_2 = f(s_2) \\
    &\HH_1 = \HH(s_1) \; , \quad \HH_2 = \HH(s_2) \; , \quad
    \Rld_1 = \Rld(s_1) \; , \quad \Rld_2 = \Rld(s_2) ...
\end{align*}

Finally, as discussed in Sec.~\ref{sec:infrared_divergence}, every time a $I_0^4$ integral appears we make the following substitution:

\begin{equation}
    I_0^4(s) \rightarrow
    \tilde{I}_0^4(s) :=
        \int_0^\infty \frac{\dd q}{2\pi^2} \,
        q^2 \, P(q) \,
        \frac{j_0(q s) - 1}{(q s)^4} \, .
\end{equation}

\subsection{Luminosity Distance TPCFs}\label{APP_LD}

\subsubsection*{LD Doppler}

\begin{equation}
    \xi^{\ndv\ndv} (s_1, s_2, y) = 
    \Jld^{\ndv\ndv}_{\alpha}
    \left[
        \Jld^{\ndv\ndv}_{\beta} 
        \left ( 
            \frac{1}{45} I_0^0(s) + 
            \frac{2}{63} I_2^0(s) +
            \frac{1}{105} I_4^0(s) 
        \right ) +
        \Jld^{\ndv\ndv}_{20} I_0^2(s)
    \right] \, ,
\end{equation}

\noindent
where

\begingroup
\allowdisplaybreaks
\begin{align}
    \Jld^{\ndv\ndv}_{\alpha} & = 
    D_1 D_2 f_1 f_2 \HH_1 \HH_2 \Rld_1 \Rld_2
    \, , \\
    \Jld^{\ndv\ndv}_{\beta} & = 
    y^2 s_1 s_2 - 2y(s_1^2 + s_2^2) + 3s_1 s_2
    \, , \\
    \Jld^{\ndv\ndv}_{20} & = \frac{1}{3} y s^2 
    \, .
\end{align}
\endgroup

\subsubsection*{LD Lensing}
\begin{align}
    \xi^{\kappa\kappa} (s_1, s_2, y) = 
    \int_0^{s_1} \dd\chi_1 \int_0^{s_2} \dd\chi_2 \;
    \Jld^{\kappa\kappa}_{\alpha}   & \left[
        \Jld^{\kappa\kappa}_{00}I^0_0(\Delta\chi) + 
        \Jld^{\kappa\kappa}_{02} I^0_2(\Delta\chi)
        \right. \nonumber
        \\
    & \left. +
        \Jld^{\kappa\kappa}_{31}I^3_1(\Delta\chi) +
        \Jld^{\kappa\kappa}_{22}I^2_2(\Delta\chi)
    \right]  \, ,
\end{align}

\noindent
where

\begingroup
\allowdisplaybreaks
\begin{align}
    \Jld^{\kappa\kappa}_{\alpha} & = 
    \frac{
        \HH_0^4 \omegaMo^2 D(\chi_1) D(\chi_2)
    }{
        s_1 s_2 a(\chi_1) a(\chi_2)
    }(\chi_1 - s_1)(\chi_2 - s_2)
    \, , \\
    \Jld^{\kappa\kappa}_{00} & = 
    -\frac{ 3 \chi_1^2 \chi_2^2}{4 \Delta\chi^4} (y^2 - 1)
    \left[
        8 y (\chi_1^2 + \chi_2^2) - 9\chi_1\chi_2y^2 - 
        7\chi_1\chi_2
    \right] 
    \, , \\
    \Jld^{\kappa\kappa}_{02} & = 
    -\frac{ 3 \chi_1^2 \chi_2^2}{2 \Delta\chi^4}(y^2 - 1)
    \left[
        4 y (\chi_1^2 + \chi_2^2) - 3 \chi_1 \chi_2 y^2 -
        5 \chi_1 \chi_2
    \right] 
    \, , \\
    \Jld^{\kappa\kappa}_{31} & = 9 y \Delta\chi^2
    \, , \\
    \Jld^{\kappa\kappa}_{22} & = 
    \frac{9 \chi_1 \chi_2}{4 \Delta\chi^4}
    \left[
        2(\chi_1^4 + \chi_2^4)(7 y^2 - 3) - 
        16  y  \chi_1 \chi_2 (\chi_1^2 + \chi_2^2)(y^2 + 1) + 
        \right.\nonumber\\
        &\left.\qquad\qquad\qquad
        \chi_1^2 \, \chi_2^2 \, (11y^4 + 14y^2 + 23) 
    \right] 
    \, .
\end{align}
\endgroup

\subsubsection*{LD Local GP}

\begin{equation}
    \xi^{\phi\phi} (s_1, s_2, y) = 
    \frac{9 \HH_0^4 \omegaMo^2 D_1 D_2 s^4 }{4 a_1 a_2}
    \left(1 + \Rld_1 + \Rld_2 + \Rld_1 \Rld_2 \right)
    \tilde{I}^4_0(s) \, .
\end{equation}

\subsubsection*{LD Integrated GP}

\begin{equation}
    \xi^{\int\!\phi\int\!\phi} (s_1, s_2, y) = \int_0^{s_1}\dd\chi_1\int_0^{s_2}\dd\chi_2 \; 
    \Jld^{\int\!\phi\int\!\phi}_{40}
    \tilde{I}^4_0(\chi) \, ,
\end{equation}

\noindent
where 

\begingroup
\allowdisplaybreaks
\begin{align}
    \Jld^{\int\!\phi\int\!\phi}_{40} = 
    \frac{
        9 \HH_0^4 \omegaMo^2 D(\chi_1) D(\chi_2) \Delta\chi^4
    }{  a(\chi_1)\, a(\chi_2)\, s_1\, s_2 } 
    &\left[
        s_2 \HH(\chi_2) \Rld_2(f(\chi_2) - 1) - 1
    \right]  \nonumber\\ 
    &\times \left[
        s_1 \HH(\chi_1) \Rld_1(f(\chi_1) - 1) - 1
    \right] \, .
\end{align}
\endgroup

\subsubsection*{LD Lensing x Doppler}

\begin{align}
    \xi^{\kappa \ndv} (s_1, s_2, y) = 
  \int_0^{s_1} \dd\chi_1 \; 
    \Jld^{\kappa \ndv}_{\alpha}  
 &   \left[
        \Jld^{\kappa \ndv}_{00} I_0^0(\Delta\chi_1) + 
        \Jld^{\kappa \ndv}_{02} I_2^0(\Delta\chi_1) 
        \right.     \nonumber
        \\
        & \left.
      +  \Jld^{\kappa \ndv}_{04} I_4^0(\Delta\chi_1) + 
        \Jld^{\kappa \ndv}_{20} I_0^2(\Delta\chi_1)
    \right]\,,
\end{align}

\noindent
where

\begingroup
\allowdisplaybreaks
\begin{align}
    \Jld^{\kappa \ndv}_{\alpha} &= 
    \HH_0^2 \omegaMo D_2 f_2 \HH_2 \Rld_2 
    \frac{D(\chi_1) (\chi_1 - s_1)}{a(\chi_1) s_1}
    \, , \\
    \Jld^{\kappa \ndv}_{00} & = 
    \frac{1}{15}
    \left[
        \chi_1^2 y + \chi_1(4 y^2 - 3) s_2 - 2 y s_2^2
    \right]
    \, , \\
    \Jld^{\kappa \ndv}_{02} & = 
    \frac{1}{42 \Delta\chi_1^2}
    \left[
        4 \chi_1^4 y + 4 \chi_1^3 (2 y^2 - 3) s_2 +
        \chi_1^2 y (11 - 23 y^2) s_2^2 
        \right.\nonumber\\
        &\left.\qquad\qquad\qquad
      +  \chi_1 (23 y^2 - 3) s_2^3 - 8 y s_2^4
    \right] 
    \, , \\
    \Jld^{\kappa \ndv}_{04} & = 
    \frac{1}{70 \Delta\chi_1^2} 
    \left[
        2\chi_1^4 y + 2 \chi_1^3 (2 y^2 - 3) s_2 -
        \chi_1^2 y (y^2 + 5) s_2^2 
        \right.\nonumber\\
        &\left.\qquad\qquad\qquad
     +    \chi_1(y^2 + 9) s_2^3 - 4 y s_2^4
    \right] 
    \, , \\
    \Jld^{\kappa \ndv}_{20} & = y \Delta\chi_1^2 
    \, .
\end{align}
\endgroup

\subsubsection*{LD Doppler X Local GP}

\begin{equation}
    \xi^{\ndv\phi} (s_1, s_2, y) = 
    \Jld^{\ndv\phi}_{31} I^3_1(s) \, ,
\end{equation}

\noindent
where

\begin{equation}
    \Jld^{\ndv\phi}_{31} = 
    \frac{3}{2 a_2} \HH_1 f_1 D_1 \Rld_1 \HH_0^2 
    \omegaMo D_2 (1 + \Rld_2)(s_1 - s_2 y) 
    \, s^2 \, .
\end{equation}

\subsubsection*{LD Doppler X Integrated GP}

\begin{equation}
    \xi^{\ndv \int\!\phi} (s_1, s_2, y) =
    \int_0^{s_2} \dd\chi_2 \;
    \Jld^{\ndv \int\!\phi}_{31} \,I_1^3(\Delta\chi_2) \, ,
\end{equation}

where

\begin{equation}
    \Jld^{\ndv \int\!\phi}_{31} =
    3 \HH_1 f_1 D_1 \HH_0^2 \omegaMo \Rld_1 
    \frac{
        D(\chi_2)(s_1 - \chi_2 y)
    }{
        a(\chi_2) s_2
    } \Delta\chi_2^2 \left[
    s_2 \Rld_2 \HH(\chi_2)(f_2 -1) - 1
    \right] \, .
\end{equation}

\subsubsection*{LD Lensing X Local GP}

\begin{equation}
    \xi^{\kappa \phi} (s_1, s_2, y) = 
    \int_0^{s_1} \dd\chi_1 
    \Jld^{\kappa \phi}_{\alpha} \left[
        \Jld^{\kappa \phi}_{31} I_1^3(\Delta\chi_1) +  
        \Jld^{\kappa \phi}_{22} I_2^2(\Delta\chi_1)
    \right]  \, ,
\end{equation}

\noindent
where

\begin{align}
    \Jld^{\kappa \phi}_{\alpha} &=
    \frac{
        9 \HH_0^4 \omegaMo^2 D_2 s_2
    }{4 a_2 s_1} (1 + \Rld_2)
    \frac{D(\chi_1)(s_1 - \chi_1)}{a(\chi_1)}
    \, , \\
    \Jld^{\kappa \phi}_{31} & = -2 y \Delta\chi_1^2 
    \, , \\
    \Jld^{\kappa \phi}_{22} & = \chi_1 s_2 (1 - y^2) 
    \, .
\end{align}

\subsubsection*{LD Lensing X Integrated GP}

\begin{equation}
    \xi^{\kappa\int\!\phi} (s_1, s_2, y) = 
    \int_0^{s_1} \dd\chi_1 \int_0^{s_2} \dd\chi_2 \;
    \Jld^{\kappa\int\!\phi}_{\alpha}\left[ 
        \Jld^{\kappa\int\!\phi}_{31} I_1^3(\Delta\chi) + 
        \Jld^{\kappa\int\!\phi}_{22} I_2^2(\Delta\chi) 
    \right] \, , 
\end{equation}

\noindent
where 

\begin{align}
    \Jld^{\kappa\int\!\phi}_{\alpha} & =
    \frac{9}{2} \HH_0^4 \omegaMo^2 
    \frac{
        D(\chi_1) D(\chi_2) \chi_2 (s_1 - \chi_1)
        }{
        s_1 a(\chi_1) a(\chi_2)
    } 
    \left[
        \HH(\chi_2)(f(\chi_2) - 1)\Rld_2 -\frac{1}{s_2} 
    \right]
    \, , \\
    \Jld^{\kappa\int\!\phi}_{31} & = -2 y \Delta\chi^2
    \, , \\
    \Jld^{\kappa\int\!\phi}_{22} & = \chi_1 \chi_2(1 - y^2) 
    \, .
\end{align}

\subsubsection*{LD Local GP X Integrated GP}
\begin{equation}
    \xi^{\phi\int\!\phi} (s_1, s_2, y) = 
    \int_0^{s_2} \dd\chi_2 \,
    \Jld^{\phi \int\!\phi}_{40} \tilde{I}^4_0(\Delta\chi_2)
    \, ,
\end{equation}

\noindent
where

\begin{equation}
    \Jld^{\phi \int\!\phi}_{40} =
    \frac{9 \HH_0^4 \omegaMo^2 D_1}{2 a_1} 
    (\Rld_1 + 1)
    \frac{D(\chi_2) \Delta\chi_2^4}{a(\chi_2)}
    \left[
        \HH(\chi_2)(f(\chi_2) - 1)\Rld_2 - \frac{1}{s_2}
    \right] \, .
\end{equation}

\subsection{Galaxy Number Counts cross Luminosity Distance perturbations TPCFs}\label{APP_GNCxLD}


\subsubsection*{Standard Newtonian GNC X Doppler LD}

\begin{align}
    \xi^{\delta \ndv}( s_1 , s_2, y ) = 
    D_1 D_2 \, \Jcr^{\delta \ndv}_{\alpha} \left[ 
        \Jcr^{\delta \ndv}_{00} I^0_{0} (s) + 
        \Jcr^{\delta \ndv}_{02} I^0_2 (s) +
        \Jcr^{\delta \ndv}_{04} I^0_4 (s) 
        \right]
\end{align}

\noindent
where

\begingroup
\allowdisplaybreaks
\begin{align}
    \Jcr^{\delta \ndv}_{\alpha} &= - f_2 \HH_2 \Rld_2
    \, , \\
    \Jcr^{\delta \ndv}_{00} &=
    \frac{1}{15} \left\{
        s_2 \left[ 5 b_1 + (2 y^2 + 1) f_1 \right] -
        y s_1 \left[ 3 f_1 + 5 b_1 \right]
    \right\} 
    \, , \\
    \Jcr^{\delta \ndv}_{02} &=
    \frac{1}{21 s^2} 
    \left\{ 
        \left[
            (y^2 + 1) f_1 + 7 b_1
        \right] s_2^3 -
        y \left[
            21 b_1 + (5 y^2 + 4) f_1 
        \right] s_1 s_2^2 
        \right.\nonumber \\
        &\left. \qquad \qquad
    +    \left[
            7 (2 y^2 + 1) b_1 + (10 y^2 - 1) f_1
        \right] s_1^2 s_2 -
        y \left[
            7 b_1 + 3 f_1
        \right] s_1^3
    \right\}
    \, , \\
    \Jcr^{\delta \ndv}_{04} &=
    \frac{f_1}{35 s^2} 
    \left[
        2 y s_1 ^3
        - 2 (y^2 + 2) s_1 ^2 s_2 +
        y (y^2 + 5) s_1 s_2^2 +
        (1 - 3 y^2) s_2 ^3
    \right] 
    \, .
\end{align}
\endgroup


\subsubsection*{Standard Newtonian GNC X Lensing LD}

\begin{equation}
    \xi^{\delta \kappa} ( s_1 , s_2, y ) =
    D_1  \int_0^{s_2}\! \dd\chi_2\; 
    \Jcr^{\delta \kappa}_{\alpha}
    \left[ 
        \Jcr^{\delta \kappa}_{00} I_0^0 ( \Delta\chi_2 ) + 
        \Jcr^{\delta \kappa}_{02} I_2^0 ( \Delta \chi_2 ) + 
        \Jcr^{\delta \kappa}_{04} I_4^0 ( \Delta \chi_2 ) 
    \right] \, ,
\end{equation}

\noindent
where

\begingroup
\allowdisplaybreaks
\begin{align}
    \Jcr^{\delta \kappa}_{\alpha} &=
    - \frac{
        \HH_0 ^2 \omegaMo D (\chi_2)
    }{
        a(\chi_2 ) s_2
    } 
    (\chi_2 - s_2 )
    \, , \\
    \Jcr^{\delta \kappa}_{00} &=
        \frac{1}{5}
        \left[
            (3 y^2 - 1) \chi_2 f_1 - y s_1(3 f_1 + 5 b_1) 
        \right] 
    \, , \\
    \Jcr^{\delta \kappa}_{02} &=
        \frac{1}{14 \Delta\chi_2^2} 
        \left\{
            4 f_1 (3 y^2 - 1) \chi_2^3 - 
            2 y 
            \left[
                (3 y^2 + 8) f_1 + 7 b_1
            \right] s_1 \chi_2^2 
            \right. \nonumber \\
            &\left.\qquad \qquad\qquad
         +   \left[
                (9 y^2 + 11) f_1 - 7 (y^2 + 3) b_1
            \right] s_1^2 \chi_2 -
            2 y \left[7 b_1 + 3 f_1 \right] s_1^3
        \right\} 
    \, , \\
    \Jcr^{\delta \kappa}_{04} &=
    \frac{f_1}{70 \Delta\chi_2^4 }
    \left\{
        (6 y^2 - 2) \chi_2^5 +
        6 y (y^2 - 3) s_1 \chi_2^4 -
        (y^4 + 12 y^2 - 21) s_1^2 \chi_2^3 
        \right.\nonumber \\
        &\left.\qquad\qquad\qquad
     +   2 y (y^2 + 3) s_1^3 \chi_2^2 -
        12 \chi_2 s_1^4 + 
        4 y s_1 ^5
    \right\} 
    \, .
\end{align}
\endgroup


\subsubsection*{Standard Newtonian GNC X Local GP LD}

\begin{align}
    \xi^{\delta \phi}( s_1 , s_2, y ) = 
    D_1 D_2 \, \Jcr^{\delta \phi}_{\alpha} \left[ 
        \Jcr^{\delta \phi}_{\beta} 
        \left(
            \frac{1}{30} I_0^0 (s) + 
            \frac{1}{21} I_2^0 (s) +
            \frac{1}{70} I_4^0 (s) 
        \right) +
        \Jcr^{\delta \phi}_{20} I_0^2 (s)
    \right] 
    \, ,
\end{align}

\noindent
where

\begingroup
\allowdisplaybreaks
\begin{align}
    \Jcr^{\delta \phi}_{\alpha} &=
    - \frac{\HH_0^2 \omegaMo}{a_2}
    (1 + \Rld_2)
    \, , \\
    \Jcr^{\delta \phi}_{\beta} &=
    f_1 \left[ 
        (3 y^2 - 1) s_2^2 - 4 y s_1 s_2 + 2 s_1^2
    \right] 
    \, , \\
    \Jcr^{\delta \phi}_{20} &= 
    - \frac{1}{2}(3 b_1 + f_1) (s_1^2 + s_2^2 - 2 y s_1 s_2)
    \, .
\end{align}
\endgroup


\subsubsection*{Standard Newtonian GNC X Integrated GP LD}

\begin{align}
\xi^{\delta \int\!\phi}( s_1 , s_2, y ) =
    D_1 \int_0^{s_2}\! & \dd\chi_2 \;  
    \Jcr^{\delta \int\!\phi}_{\alpha}
    \left[ 
        \Jcr^{\delta \int\!\phi}_{20} I_0^2 ( \Delta\chi_2 ) 
        \right.  \nonumber\\
        & \left. 
    +    \Jcr^{\delta \int\!\phi}_{\beta}
            \left(
                \frac{1}{15} I_0^0 ( \Delta\chi_2 ) + 
                \frac{1}{21} I_2^0 ( \Delta\chi_2 ) +
                \frac{1}{35} I_4^0 ( \Delta\chi_2 )
            \right) 
    \right] \, , 
\end{align}

\noindent
where

\begingroup
\allowdisplaybreaks
\begin{align}
    \Jcr^{\delta \int\!\phi}_{\alpha} &=
    - \frac{\HH_0^2 \omegaMo D(\chi_2)}{3 a(\chi_2) s_2} 
    \left[ 
        s_2 \Rld_2 \HH(\chi_2) ( f(\chi_2) - 1) - 1
    \right] 
    \, , \\
    \Jcr^{\delta \int\!\phi}_{\beta} &=
    f_1 \left[ 
        (3 y^2 - 1) \chi_2^2 - 4 y s_1 \chi_2 + 2 s_1^2
    \right] 
    \, , \\
    \Jcr^{\delta \int\!\phi}_{20} &=
    - \Delta\chi_2^2 ( 3 b_1 + f_1)
    \, .
\end{align}
\endgroup

\subsubsection*{Doppler GNC X Doppler LD}

\begin{align}
    \xi^{\ndv \ndv}( s_1, s_2 , y ) =
    D_1 D_2 \, \Jcr_{\alpha}^{\ndv \ndv}
    \left[  
        \Jcr_{\beta}^{\ndv \ndv}
        \left( 
            \frac{1}{45} I_0^0(s) +
            \frac{2}{63} I_2^0 (s) + 
            \frac{1}{105} I_4^0(s)
        \right) +
        \Jcr^{\ndv \ndv}_{20} I_0^2 (s)
    \right] 
    \, ,
\end{align}

\noindent
where

\begingroup
\allowdisplaybreaks
\begin{align}
    \Jcr_{\alpha}^{\ndv \ndv} &= 
    - f_1 f_2\HH_1 \HH_2 \Rgnc_1 \Rld_2
    \, , \\
    \Jcr_{\beta}^{\ndv \ndv} &= 
    y^2 s_1 s_2 - 2 y (s_1^2 + s_2^2) + 3 s_1 s_2
    \, , \\
    \Jcr_{20}^{\ndv \ndv}  &=
    \frac{1}{3} y s^2
    \, .
\end{align}
\endgroup


\subsubsection*{Doppler GNC X Lensing LD}

\begin{align}
    \xi^{\kappa \ndv} (s_1, s_2, y) = 
    D_1 \int_0^{s_2} \! \dd\chi_2 \; 
    \Jcr^{\kappa \ndv}_{\alpha} & \left[
        \Jcr^{\kappa \ndv}_{00} I_0^0(\Delta\chi_2) + 
        \Jcr^{\kappa \ndv}_{02} I_2^0(\Delta\chi_2) 
     \nonumber \right.   \\
    & \left. + 
        \Jcr^{\kappa \ndv}_{04} I_4^0(\Delta\chi_2) + 
        \Jcr^{\kappa \ndv}_{20} I_0^2(\Delta\chi_2)
    \right]\,,
\end{align}

\noindent
where

\begingroup
\allowdisplaybreaks
\begin{align}
    \Jcr^{\kappa \ndv}_{\alpha} &= 
    - \HH_0^2 \omegaMo f_1 \HH_1 \Rgnc_1 
    \frac{D(\chi_2) (\chi_2 - s_2)}{a(\chi_2) s_2}
    \, , \\
    \Jcr^{\kappa \ndv}_{00} & = 
    \frac{1}{15}
    \left[
        \chi_2^2 y + \chi_2(4 y^2 - 3) s_1 - 2 y s_1^2
    \right]
    \, , \\
    \Jcr^{\kappa \ndv}_{02} & = 
    \frac{1}{42 \Delta\chi_2^2}
    \left[
        4 \chi_2^4 y + 4 \chi_2^3 (2 y^2 - 3) s_1 +
        \chi_2^2 y (11 - 23 y^2) s_1^2 
        \right.\nonumber\\
        &\left.\qquad\qquad\qquad
     +   \chi_2 (23 y^2 - 3) s_1^3 - 8 y s_1^4
    \right] 
    \, , \\
    \Jcr^{\kappa \ndv}_{04} & = 
    \frac{1}{70 \Delta\chi_2^2} 
    \left[
        2\chi_2^4 y + 2 \chi_2^3 (2 y^2 - 3) s_1 -
        \chi_2^2 y (y^2 + 5) s_1^2 
        \right.\nonumber\\
        &\left.\qquad\qquad\qquad
   +      \chi_2 (y^2 + 9) s_1^3 - 4 y s_1^4
    \right] 
    \, , \\
    \Jcr^{\kappa \ndv}_{20} & = y \Delta\chi_2^2 
    \, .
\end{align}
\endgroup


\subsubsection*{Doppler GNC X Local GP LC}

\begin{equation}
    \xi^{\ndv\phi} (s_1, s_2, y) = 
    D_1 \, D_2 \, \Jcr^{\ndv\phi}_{31} I^3_1(s) \, ,
\end{equation}

\noindent
where

\begin{equation}
    \Jcr^{\ndv\phi}_{31} = 
    - \frac{3}{2 a_2} f_1 \HH_1 \Rgnc_1 \HH_0^2 
    \omegaMo (1 + \Rld_2)(y s_2 - s_1) 
    \, s^2 \, .
\end{equation}


\subsubsection*{Doppler GNC X Integrated GP LD}

\begin{equation}
    \xi^{\ndv \int\!\phi} (s_1, s_2, y) =
    D_1 \int_0^{s_2} \! \dd\chi_2 \;
    \Jcr^{\ndv \int\!\phi}_{31} \,I_1^3(\Delta\chi_2) \, ,
\end{equation}

\noindent
where

\begin{equation}
    \Jcr^{\ndv \int\!\phi}_{31} =
    3 f_1 \HH_1 \Rgnc_1 \HH_0^2 \omegaMo 
    \frac{
        D(\chi_2)(\chi_2 y - s_1)
    }{
        a(\chi_2) s_2
    } \Delta\chi_2^2 \left[
    s_2 \Rld_2 \HH(\chi_2)(f_2 -1) - 1 
    \right] \, .
\end{equation}

\subsubsection*{Lensing GNC X Doppler LD}

\begin{align}
    \xi^{\kappa \ndv} (s_1, s_2, y) = 
    D_2 \int_0^{s_1} \! \dd\chi_1 \; 
    \Jcr^{\kappa \ndv}_{\alpha} & \left[
        \Jcr^{\kappa \ndv}_{00} I_0^0(\Delta\chi_1) + 
        \Jcr^{\kappa \ndv}_{02} I_2^0(\Delta\chi_1) 
        \nonumber \right.
         \\
    & \left.
        + 
        \Jcr^{\kappa \ndv}_{04} I_4^0(\Delta\chi_1) + 
        \Jcr^{\kappa \ndv}_{20} I_0^2(\Delta\chi_1)
    \right]\,,
\end{align}

\noindent
where

\begingroup
\allowdisplaybreaks
\begin{align}
    \Jcr^{\kappa \ndv}_{\alpha} &= 
    - \HH_0^2 \omegaMo f_2 \HH_2 \Rld_2 
    \frac{D(\chi_1) (\chi_1 - s_1)}{a(\chi_1) s_1}
    (5 \mbs{1} - 2 )
    \, , \\
    \Jcr^{\kappa \ndv}_{00} & = 
    \frac{1}{15}
    \left[
        \chi_1^2 y + \chi_1(4 y^2 - 3) s_2 - 2 y s_2^2
    \right]
    \, , \\
    \Jcr^{\kappa \ndv}_{02} & = 
    \frac{1}{42 \Delta\chi_1^2}
    \left[
        4 \chi_1^4 y + 4 \chi_1^3 (2 y^2 - 3) s_2 +
        \chi_1^2 y (11 - 23 y^2) s_2^2 
        \right.\nonumber \\
        &\left.\qquad\qquad\qquad
    +    \chi_1 (23 y^2 - 3) s_2^3 - 8 y s_2^4
    \right] 
    \, , \\
    \Jcr^{\kappa \ndv}_{04} & = 
    \frac{1}{70 \Delta\chi_1^2} 
    \left[
        2\chi_1^4 y + 2 \chi_1^3 (2 y^2 - 3) s_2 -
        \chi_1^2 y (y^2 + 5) s_2^2 
        \right. \nonumber \\
        &\left.\qquad\qquad\qquad
   +      \chi_1(y^2 + 9) s_2^3 - 4 y s_2^4
    \right] 
    \, , \\
    \Jcr^{\kappa \ndv}_{20} & = y \Delta\chi_1^2 
    \, .
\end{align}
\endgroup


\subsubsection*{Lensing GNC x Lensing LD}

\begin{align}
    \xi^{\kappa\kappa} (s_1, s_2, y) = 
    \int_0^{s_1} \! \dd\chi_1 \int_0^{s_2} \! \dd\chi_2\;  
    \Jcr^{\kappa\kappa}_{\alpha}
    &\left[
        \Jcr^{\kappa\kappa}_{00} I_0^0(\Delta\chi) + 
        \Jcr^{\kappa\kappa}_{02} I_2^0(\Delta\chi) 
        \right.\nonumber\\
        &\left.
      +  \Jcr^{\kappa\kappa}_{31} I_1^3(\Delta\chi) +
        \Jcr^{\kappa\kappa}_{22} I_2^2(\Delta\chi)
    \right]  \, , 
\end{align}

\noindent
where

\begingroup
\allowdisplaybreaks
\begin{flalign}
    \Jcr^{\kappa\kappa}_{\alpha} & = 
    - \frac{
        \HH_0^4 \omegaMo^2 D(\chi_1) D(\chi_2) 
    }{
        s_1 s_2 a(\chi_1) a(\chi_2)}
    (\chi_1 - s_1)(\chi_2 - s_2)
    (5 \mbs{1} - 2)
    \, , \\
    \Jcr^{\kappa\kappa}_{00} & = 
    -\frac{ 3 \chi_1^2 \chi_2^2}{4 \Delta\chi^4} (y^2 - 1)
    \left[
        8 y (\chi_1^2 + \chi_2^2) - 9\chi_1\chi_2y^2 - 
        7\chi_1\chi_2
    \right] 
    \, , \\
    \Jcr^{\kappa\kappa}_{02} & = 
    -\frac{ 3 \chi_1^2 \chi_2^2}{2 \Delta\chi^4}(y^2 - 1)
    \left[
        4 y (\chi_1^2 + \chi_2^2) - 3 \chi_1 \chi_2 y^2 -
        5 \chi_1 \chi_2
    \right] 
    \, , \\
    \Jcr^{\kappa\kappa}_{31} & = 9 y \Delta\chi^2 
    \, , \\
    \Jcr^{\kappa\kappa}_{22} & = 
    \frac{9 \chi_1 \chi_2}{4 \Delta\chi^4}
    \left[
        2(\chi_1^4 + \chi_2^4)(7 y^2 - 3) - 
        16 y \chi_1 \chi_2 (\chi_1^2 + \chi_2^2)(y^2 + 1) 
        \right.  \nonumber \\
        &\left.\qquad\qquad\qquad
    +    \chi_1^2 \chi_2^2 (11y^4 + 14y^2 + 23) 
    \right] 
    \, .
\end{flalign}
\endgroup


\subsubsection*{Lensing GNC X Local GP LD}

\begin{equation}
    \xi^{\kappa \phi} (s_1, s_2, y) = 
    D_2 \int_0^{s_1} \! \dd\chi_1 \;
    \Jcr^{\kappa \phi}_{\alpha} \left[
        \Jcr^{\kappa \phi}_{31} I_1^3(\Delta\chi_1) +  
        \Jcr^{\kappa \phi}_{22} I_2^2(\Delta\chi_1)
    \right]  \, ,
\end{equation}

\noindent
where

\begin{align}
    \Jcr^{\kappa \phi}_{\alpha} &=
    - \frac{
        9 \HH_0^4 \omegaMo^2 s_2 D(\chi_1)(s_1 - \chi_1)
    }{
        4 a_2 s_1a(\chi_1)
    } 
    (1 + \Rld_2)
    (5 \mbs{1} - 2)
    \, , \\
    \Jcr^{\kappa \phi}_{31} & = -2 y \Delta\chi_1^2 
    \, , \\
    \Jcr^{\kappa \phi}_{22} & = \chi_1 s_2 (1 - y^2) 
    \, .
\end{align}


\subsubsection*{Lensing GNC X Integrated GP LD}

\begin{equation}
    \xi^{\kappa\int\!\phi} (s_1, s_2, y) = 
    \int_0^{s_1} \! \dd\chi_1 \int_0^{s_2} \! \dd\chi_2 \;
    \Jcr^{\kappa\int\!\phi}_{\alpha}\left[ 
        \Jcr^{\kappa\int\!\phi}_{31} I_1^3(\Delta\chi) + 
        \Jcr^{\kappa\int\!\phi}_{22} I_2^2(\Delta\chi) 
    \right] \, , 
\end{equation}

\noindent
where 

\begin{align}
    \Jcr^{\kappa\int\!\phi}_{\alpha} & =
    \frac{9}{2} \HH_0^4 \omegaMo^2 
    \frac{
        D(\chi_1) D(\chi_2) \chi_2 (s_1 - \chi_1)
        }{
        s_1 s_2 a(\chi_1) a(\chi_2)
    } (5 \mbs{1} - 2 )
    \left[
        s_2 \Rld_2 \HH(\chi_2) (f(\chi_2) - 1) - 1
    \right]
    \, , \\
    \Jcr^{\kappa\int\!\phi}_{31} & = -2 y \Delta\chi^2
    \, , \\
    \Jcr^{\kappa\int\!\phi}_{22} & = \chi_1 \chi_2(1 - y^2) 
    \, .
\end{align}

\subsubsection*{Local GP GNC X Doppler LD}

\begin{align}
    \xi^{\phi \ndv} ( s_1 , s_2, y ) = 
    D_1 D_2 \, \Jcr^{\phi \ndv}_{\alpha}
    \left[ 
        \frac{1}{90} I_0^0 (s) +
        \frac{1}{63} I_2^0 (s) + 
        \frac{1}{210} I_4^0 (s) +
        \frac{1}{6} I_0^2 (s) 
    \right]  ,
\end{align}

\noindent
where

\begingroup
\allowdisplaybreaks
\begin{align}
    \Jcr^{\phi \ndv}_{\alpha} &=
    - \frac{f_2 \HH_2 \Rld_2 s^2}{a_1} (y s_1 - s_2) 
   \nonumber  \\
    & \qquad
   \times \left[ 
        2 f_1 a_1 \HH_1^2 (\fevos{1} - 3) + 
        3 \HH_0^2 \omegaMo (f_1 + \Rgnc_1 + 5 \mbs{1} - 2)
    \right]
     \, .
\end{align}
\endgroup


\subsubsection*{Local GP GNC X Lensing LD}

\begin{align}
    \xi^{\phi \kappa} ( s_1 , s_2, y ) &= 
    D_1 \int_0^{s_2}\! \dd\chi_2 \; 
    \Jcr^{\phi \kappa}_{\alpha}\left[ 
        \Jcr^{\phi \kappa}_{20} I_0^2 ( \Delta \chi_2 ) 
        \right.\nonumber \\
        &\left.\qquad
     +   \Jcr^{\phi \kappa}_{\beta}
        \left(
            \frac{1}{60} I_0^0 ( \Delta \chi_2 ) +
            \frac{1}{42} I_2^0 ( \Delta \chi_2 ) +
            \frac{1}{140} I_4^0 ( \Delta \chi_2 ) 
        \right)
    \right] \, ,
\end{align}

\noindent
where

\begingroup
\allowdisplaybreaks
\begin{align}
    \Jcr^{\phi \kappa}_{\alpha}  &= 
    - \frac{\HH_0^2 \omegaMo s_1 D(\chi_2)}{ a_1 s_2 a(\chi_2)}
    (\chi_2 - s_2) 
    \\
    &\qquad\qquad
    \times\left[
       2 f_1 a_1 \HH_1^2 (\fevos{1} - 3) + 
       3 \HH_0^2 \omegaMo (f_1 + \Rgnc_1 + 5 \mbs{1} - 2)
    \right] \nonumber
    \, , \\
    \Jcr^{\phi \kappa}_{\beta} &=
    2 y \chi_2^2 - \chi_2 s_1 (y^2 + 3) + 2 y s_1^2
    \, ,\\
    \Jcr^{\phi \kappa}_{20} &= \frac{1}{2} y \Delta\chi_2^2\, .
\end{align}
\endgroup


\subsubsection*{Local GP GNC X Local GP LD}

\begin{align}\label{eq:TPCF_phiGNC_phiLD}
    \xi^{\phi\phi}( s_1 , s_2, y ) &= 
    D_1 D_2 \, \Jcr_{40}^{\phi\phi}( s_1, s_2 ) \tilde{I}_0^4 (s)  
    \, ,
\end{align}

\noindent
where

\begingroup
\allowdisplaybreaks
\begin{align}
    \Jcr_{40}^{\phi\phi} (s_1, s_2)  &= 
    - \frac{3 \HH_0^2 \omegaMo s^4}{4 a_1 a_2}
    (1 + \Rld_2)
    \left[
        2 a_1 f_1(\fevos{1}-3) \HH_1^2 + 
        3 \HH_0^2 \omegaMo (f_1 + \Rgnc_1 + 5 \mbs{1} - 2)
    \right] 
    \, .
\end{align}
\endgroup


\subsubsection*{Local GP GNC X Integrated GP LD}

\begin{equation}
    \xi^{\phi \int\!\phi} ( s_1 , s_2, y ) = 
    D_1 \int_0^{s_2}\! \dd\chi_2 \;
    \Jcr^{\phi \int\!\phi}_{40} I_0^4 ( \Delta\chi_2 )  
    \, ,
\end{equation}

\noindent
where

\begingroup
\allowdisplaybreaks
\begin{align}
    \Jcr^{\phi \int\!\phi}_{40} &=
    - \frac{
        3 \Delta\chi_2^4 \HH_0^2 \omegaMo D(\chi_2) 
    }{
        2 s_2 a(\chi_2) a_1
    } \left[
        s_2 \HH(\chi_2) \Rld_2 ( f(\chi_2) - 1 ) - 1
    \right] \nonumber
  \\
    &\qquad\qquad\qquad
     \times \left[
        2 a_1 f_1 \HH_1^2 (\fevos{1} - 3) +
        3 \HH_0^2 \omegaMo (f_1 + \Rgnc_1 + 5 \mbs{1} - 2)
    \right]
    \, . 
\end{align}
\endgroup

\subsubsection*{Integrated GP GNC X Doppler LD}

\begin{equation}
    \xi^{\int\!\phi \, \ndv} (s_1, s_2, y) =
    D_2 \int_0^{s_1} \dd\chi_1 \;
    \Jcr^{\int\!\phi \, \ndv}_{31} \,I_1^3(\Delta\chi_1) \, ,
\end{equation}

\noindent
where

\begin{equation}
    \Jcr^{\int\!\phi \, \ndv}_{31} =
    - 3 f_2 \HH_2 \Rld_2  \HH_0^2 \omegaMo 
    \frac{
        D(\chi_1)(s_2 - \chi_1 y)
    }{
        a(\chi_1) s_1
    } \Delta\chi_1^2 \left[
    s_1 \Rgnc_1 \HH(\chi_1)(f_1 - 1) - 5 \mbs{1} + 2
    \right] \, .
\end{equation}


\subsubsection*{Integrated GP GNC X Lensing LD}

\begin{equation}\label{eq:TPCF_intphiGNC_kappaLD}
    \xi^{\int\!\phi \, \kappa} ( s_1 , s_2, y ) = 
    \int_0^{s_1}\! \dd\chi_1 \int_0^{s_2}\! \dd\chi_2 \;
    \Jcr_{\alpha}^{\int\!\phi \, \kappa} 
    \left[ 
        \Jcr_{31}^{\int\!\phi \, \kappa} I_1^3 ( \Delta \chi ) +
        \Jcr_{22}^{\int\!\phi \, \kappa} I_2^2 ( \Delta \chi ) 
     \right] \, ,
\end{equation}

\noindent
where

\begingroup
\allowdisplaybreaks
\begin{align}
    \Jcr_{\alpha}^{\int\!\phi \, \kappa} &=
    - \frac{
        9 \chi_1 \HH_0^4 \omegaMo^2 D(\chi_1) D(\chi_2) 
    }{
        a(\chi_1)  a(\chi_2) s_1  s_2
    }
    (\chi_2 - s_2)
    \left[
        \HH(\chi_1) \Rgnc_1 s_1 (f(\chi_1) - 1) - 5 \mbs{1} + 2
    \right]
    \, , \\
    \Jcr_{31}^{\int\!\phi \, \kappa} &=  y \Delta\chi^2
    \, , \\
    \Jcr_{22}^{\int\!\phi \, \kappa} &= 
    \frac{1}{2} (y^2 - 1) \chi_1 \chi_2 
    \, .
\end{align}
\endgroup


\subsubsection*{Integrated GP GNC X Local GP LD}
\begin{equation}
    \xi^{\int\!\phi \, \phi} (s_1, s_2, y) = 
    D_2 \, \int_0^{s_1} \dd\chi_1 \;
    \Jcr^{\int\!\phi \, \phi}_{40} \tilde{I}^4_0(\Delta\chi_1)
    \, ,
\end{equation}

\noindent
where

\begin{equation}
    \Jcr^{\int\!\phi \, \phi}_{40} =
    - \frac{9 \HH_0^4 \omegaMo^2 D(\chi_1) }{2 a_2 s_1 a(\chi_1)} 
    \Delta\chi_1^4 (1 + \Rld_2)
    \left[
        s_1 \HH(\chi_1) \Rgnc_1 (f(\chi_1) - 1) - 5 \mbs{1} + 2
    \right] \, .
\end{equation}


\subsubsection*{Integrated GP GNC X Integrated GP LD}

\begin{equation}
    \xi^{\int\!\phi \, \int \!\phi }( s_1 , s_2, y ) = \int_0^{s_1}\! \dd\chi_1  \int_0^{s_2}\! \dd\chi_2 \;  
    \Jcr^{\int \!\phi \int \!\phi}_{40} 
    \tilde{I}_0^4 ( \Delta\chi) \, ,
\end{equation}

\noindent
where

\begin{align}
\Jcr^{\int \!\phi\int \!\phi}_{40} =
    - \frac{
        9 \Delta\chi ^4 \HH_0^4 \omegaMo^2 D(\chi_1) D(\chi_2)
    }{
        a(\chi_1) a(\chi_2) s_1 s_2
    }
    &\left[
        s_1 (f(\chi_1) - 1) \HH(\chi_1) \Rgnc_1 - 5 \mbs{1} + 2
    \right] 
    \nonumber \\
    &\times\left[
        s_2 (f(\chi_2) - 1) \HH(\chi_2) \Rld_2 - 1
    \right]
    \, .
\end{align}

\section{Definition of the distance perturbation}

\label{app:mean_distance}
In this Appendix we want to discuss how luminosity distance fluctuations are constructed in peculiar velocity surveys.
We can define the luminosity (or angular diameter) distance perturbations as
\begin{eqnarray}  \label{eq:def_DeltaDL}
\Delta D_L \left( z , \bn \right) \equiv \frac{D_L \left( z, \bn \right) - \langle D_L \left( z, \bn \right) \rangle }{\langle D_L \left( z, \bn \right) \rangle}\,,
\end{eqnarray}
where $\langle .. \rangle$ denotes the angular average at fixed observed redshift.
To relate this measurement with some theoretical prediction we need to introduce a theoretical luminosity distance $D^{\rm th}_L$ such that
\begin{equation}
    D_L \left( z , \bn \right)  = D^{\rm th}_L \left( \bar z , \bar \bn \right) \, ,
\end{equation}
where $\bar z $ and $\bar \bn$ are the cosmological background redshift and the unperturbed direction, respectively.
Following the Ergodic theorem, we replace the angular average with the ensemble average of the theoretical luminosity distance obtaining
\begin{equation}
    \langle D_L \left( z, \bn \right) \rangle = \bar D^{\rm th}_L \left( \bar z \right) \, .
\end{equation}
In order to derive the luminosity distance perturbation $\Delta D_L$ shown in \eqref{eq:fluctuations_ld}, we need to compute $D^{\rm th}_L \left( \bar z , \bar \bn \right)$ to first order. However since we can only measure $z$, we need to take into account the difference in the background luminosity density between the observed redshift $z$ and the background one $\bar z$. This implies that the luminosity distance perturbation can be written as
\begin{eqnarray} \nonumber
    \Delta D_L \left( z , \bn \right) &=& \frac{D^{\rm th}_L \left( \bar z , \bar \bn \right) -\bar D^{\rm th}_L \left( \bar z \right) }{\bar D^{\rm th}_L \left( \bar z \right)}
       \nonumber \\
    &\simeq&
    \frac{D^{\rm th}_L \left(  z ,  \bn \right) - \frac{dD^{\rm th}_L}{dz} \delta z -  D^{\rm th}_L \left( z \right) + \frac{d \bar D^{\rm th}_L }{dz} \delta z } { D^{\rm th}_L \left( z \right) - \frac{d \bar D^{\rm th}_L }{dz} \delta z }
    \simeq\frac{\delta D_L^{\rm th} \left( z, \bn \right) }{D^{\rm th}_L \left( z \right)}\, .
\end{eqnarray}
In this approach we have therefore assumed that the number of sourced is large enough to compute a meaningful angular average at fixed observed redshift.

Due to the small sample size, this is not what has been done so far in peculiar velocity surveys.
In current analyses, see e.g.~\cite{SDSS:2014iwm}, the measured luminosity distance is compared with the background distance in a fiducial model evaluated at a reconstructed background redshift $\bar z$.
Therefore the luminosity distance perturbation is defined \emph{object by object} as
\begin{equation}
    \Delta d_L \left( z , \bn \right) \equiv \frac{D_L \left( z, \bn \right) - \bar D_L^{\rm fid}\left( \bar z \right)}{\bar D_L^{\rm fid}\left( \bar z \right)}\,,
\end{equation}
and it is related to $\Delta D_L \left( z , \bn \right)$, defined in \eqref{eq:def_DeltaDL}, via
\begin{eqnarray}
      \Delta d_L \left( z , \bn \right) &=& \frac{ \bar D^{\rm th}_L \left( \bar z \right) \left( 1 + \Delta D_L \left( z , \bn \right) \right)}{\bar D_L^{\rm fid}\left( \bar z \right)} -1
      \nonumber \\
&\simeq&
\frac{\bar D_L^{\rm th} \left(  z \right)}{\bar D_L^{\rm fid}  \left( z \right)}
\left( 1+\Delta D_L \left( z , \bn \right) -\left( \frac{d }{dz} \ln\frac{ \bar D_L^{\rm th}}{ \bar D_L^{\rm fid}} \right) \delta z \right)   -1 \,.
\end{eqnarray}
Therefore we find that as long as the difference between the fiducial model and the true underlying cosmology is small, the difference between the equation above and Eq.~\ref{eq:def_DeltaDL} is much smaller than the characteristic size of the linear perturbations, modulo an over-all rescaling due to the assumption of a fiducial cosmology that can be easily be included in the modeling.

Let us also note, as already discussed in~\cite{Calcino:2016jpu}, that using directly the observed redshift $z$ in the fiducial model would introduce a large error.
Indeed by defining the luminosity distance perturbation as
\begin{equation}
    \Delta \tilde d_L \left( z , \bn \right) \equiv \frac{D_L \left( z, \bn \right) - \bar D_L^{\rm fid}\left(  z \right)}{\bar D_L^{\rm fid}\left(  z \right)}\,,
\end{equation}
we find that the relation with $\Delta D_L$ is given by
\begin{eqnarray}
        \Delta \tilde d_L \left( z , \bn \right) &=& \frac{ \bar D^{\rm th}_L \left( \bar z \right) \left( 1 + \Delta D_L \left( z , \bn \right) \right)}{\bar D_L^{\rm fid}\left(  z \right)} -1
      \nonumber \\
&\simeq&
\frac{\bar D_L^{\rm th} \left(  z \right)}{\bar D_L^{\rm fid}  \left( z \right)}
\left( 1+\Delta D_L \left( z , \bn \right) -\left( \frac{d }{dz} \ln \bar D_L^{\rm th}\right) \delta z \right)   -1 \, .
\end{eqnarray}
Now the prefactor of the redshift perturbation $\delta z$ is non-negligible, ranging from  $20$ at $z=0.05$ to $5$ at $z=0.2$, and it does not vanish even by considering the correct fiducial model.

\section{Multipoles covariance}
\label{app:Covariance_derivation}

Let us start from the definition of the covariance given in the main text, \eqref{eq:covariance_definition}, and expand every power spectrum in Legendre polynomials
\begin{equation}
\begin{split}
     \mathcal{C}^{(XYWZ)}(\mathbf{k}_1, \mathbf{k}_2) & \equiv
    P^{(XW)}(\mathbf{k}_1)\delta_D^{(3)}(\mathbf{k}_1 - \mathbf{k}_2)P^{(YZ)}(\mathbf{-k}_1)\delta_D^{(3)}(\mathbf{k}_1 - \mathbf{k}_2) \\
    & +  P^{(XZ)}(\mathbf{k}_1)\delta_D^{(3)}(\mathbf{k}_1 + \mathbf{k}_2)P^{(YW)}(\mathbf{-k}_1)\delta_D^{(3)}(\mathbf{k}_1 + \mathbf{k}_2)  = \\
    & = \sum_{\ell_1, \ell_2} [P_{\ell_1}^{(XW)}(k_1)\mathcal{L}_{\ell_1}(\mu_1)P_{\ell_2}^{(YZ)}(k_1)\mathcal{L}_{\ell_2}(-\mu_1)\delta_D^{(3)}(\mathbf{k}_1 - \mathbf{k}_2)\delta_D^{(3)}(\mathbf{k}_1 - \mathbf{k}_2) \\
    & + P_{\ell_1}^{(XZ)}(k_1)\mathcal{L}_{\ell_1}(\mu_1)P_{\ell_2}^{(YW)}(k_1)\mathcal{L}_{\ell_2}(-\mu_1)\delta_D^{(3)}(\mathbf{k}_1 + \mathbf{k}_2)\delta_D^{(3)}(\mathbf{k}_1 + \mathbf{k}_2)]\,.
\end{split}
\end{equation}
We compute the multipoles of the covariance as follows
\begin{equation}
\begin{split}
    \mathcal{ C}^{(XYWZ)}_{L_1 L_2}& = \frac{2L_1+1}{2}\frac{2L_2+1}{2} \sum_{\ell_1, \ell_2} \bigg[ P_{\ell_1}^{(XW)}(k_1) P_{\ell_2}^{(YZ)}(k_1) \int d\mu_1 d\mu_2  \mathcal{L}_{\ell_1}(\mu_1)\mathcal{L}_{\ell_2}(-\mu_1) \\
    & \times \mathcal{L}_{L_1}(\mu_1) \mathcal{L}_{L_2}(\mu_2) \delta_D^{(3)}(\mathbf{k}_1 - \mathbf{k}_2)\delta_D^{(3)}(\mathbf{k}_1 - \mathbf{k}_2) + P_{\ell_1}^{(XZ)}(k_1) P_{\ell_2}^{(YW)}(k_1) \\
    & \times \int d\mu_1 d\mu_2  \mathcal{L}_{\ell_1}(\mu_1)\mathcal{L}_{\ell_2}(-\mu_1) \mathcal{L}_{L_1}(\mu_1) \mathcal{L}_{L_2}(\mu_2) \delta_D^{(3)}(\mathbf{k}_1 + \mathbf{k}_2)\delta_D^{(3)}(\mathbf{k}_1 + \mathbf{k}_2) \bigg]\,.
\end{split}
\end{equation}
We can now split the 3D Dirac deltas into a radial and an angular part; the radial part can be taken out from the integrals while the angular part can be used to compute the $\mu_2$ integral, leading to (renaming $\mu_1 \rightarrow \mu$)
\begin{equation}
\begin{split}
    \mathcal{ C}^{(XYWZ)}_{L_1 L_2} &= \frac{2L_1+1}{2}\frac{2L_2+1}{2} \frac{\delta_D(k_1-k_2)}{k_1^2} \delta_D^{(3)} (0)
    \\
    &
    \sum_{\ell_1, \ell_2} \bigg[ P_{\ell_1}^{(XW)}(k_1) P_{\ell_2}^{(YZ)}(k_1) \int d\mu \mathcal{L}_{\ell_1}(\mu)\mathcal{L}_{\ell_2}(-\mu) \mathcal{L}_{L_1}(\mu) \mathcal{L}_{L_2}(\mu)\\
    &   + P_{\ell_1}^{(XZ)}(k_1) P_{\ell_2}^{(YW)}(k_1) \int d\mu \mathcal{L}_{\ell_1}(\mu)\mathcal{L}_{\ell_2}(-\mu) \mathcal{L}_{L_1}(\mu) \mathcal{L}_{L_2}(-\mu) \bigg]\,.
     \\
   &= \frac{\delta_D(k_1-k_2)}{k_1^2} \delta_D^{(3)} (0) {C}^{(XYWZ)}_{L_1 L_2}
\end{split}
\end{equation}
Let us focus on the two $\mu$ integrals; first of all, we can use $\mathcal{L}_\ell(-\mu) = (-1)^\ell \mathcal{L}_\ell(\mu)$ to have all the Legendre polynomials evaluated at the same angle. The remaining integral to be solved is then
\begin{equation}
    I = \int d\mu \mathcal{L}_{\ell_1}(\mu)\mathcal{L}_{\ell_2}(\mu) \mathcal{L}_{L_1}(\mu) \mathcal{L}_{L_2}(\mu)
\end{equation}
in both cases. Recall the following relation regarding the product of two Legendre polynomials
\begin{equation}
    \mathcal{L}_\ell(\mu) \mathcal{L}_L(\mu) = \sum_{n=|\ell - L|}^{\ell+L}\bigg(
    \begin{matrix}
    L & \ell & n \\
    0 & 0 & 0
    \end{matrix}
    \bigg)^2(2n+1)\mathcal{L}_n(\mu)\,,
\end{equation}
which we can use to transform the $\mu$ integral from an integral of the product of 4 Legendre polynomials to an integral of the product of 2 Legendre polynomials
\begin{equation}
    I = \sum_{n_1, n_2} \bigg(
    \begin{matrix}
    \ell_1 & L_1 & n_1 \\
    0 & 0 & 0
    \end{matrix}
    \bigg)^2 (2n_1+1)\bigg(
    \begin{matrix}
    \ell_2 & L_2 & n_2 \\
    0 & 0 & 0
    \end{matrix}
    \bigg)^2 (2n_2+1) \int d\mu \mathcal{L}_{n_1}(\mu) \mathcal{L}_{n_2}(\mu)\,.
\end{equation}
Finally we have, from the orthogonality of Legendre polynomials
\begin{equation}
    \int d\mu \mathcal{L}_{n_1}(\mu) \mathcal{L}_{n_2}(\mu) = \frac{2}{2n_1+1}\delta_{n_1 n_2}\,,
\end{equation}
so that the sum over $n_2$ can be computed using the Kronecker delta and the expression for the multipoles of the covariance matrix reads (reinserting the $(-1)^\ell$ factors in the correct places and renaming $n_1 \rightarrow n$)
\begin{equation}
\begin{split}
    C ^{(XYWZ)}_{L_1 L_2} = & \frac{2L_1+1}{2}(2L_2+1) \sum_{\ell_1,\ell_2, n}(-1)^{\ell_2}\bigg(
    \begin{matrix}
    \ell_1 & L_1 & n \\
    0 & 0 & 0
    \end{matrix}
    \bigg)^2 \bigg(
    \begin{matrix}
    \ell_2 & L_2 & n \\
    0 & 0 & 0
    \end{matrix}
    \bigg)^2 \\
    & \times  (2n+1) [P_{\ell_1}^{(XW)}(k_1) P_{\ell_2}^{(YZ)}(k_1) + (-1)^{L_2}P_{\ell_1}^{(XZ)}(k_1) P_{\ell_2}^{(YW)}(k_1)]\,,
\end{split}
\end{equation}
which leads to \eqref{eq:cov}.

\section{The multipoles of the galaxy number counts power spectrum}
\begin{figure}
    \centering
\includegraphics[width=0.32\textwidth]{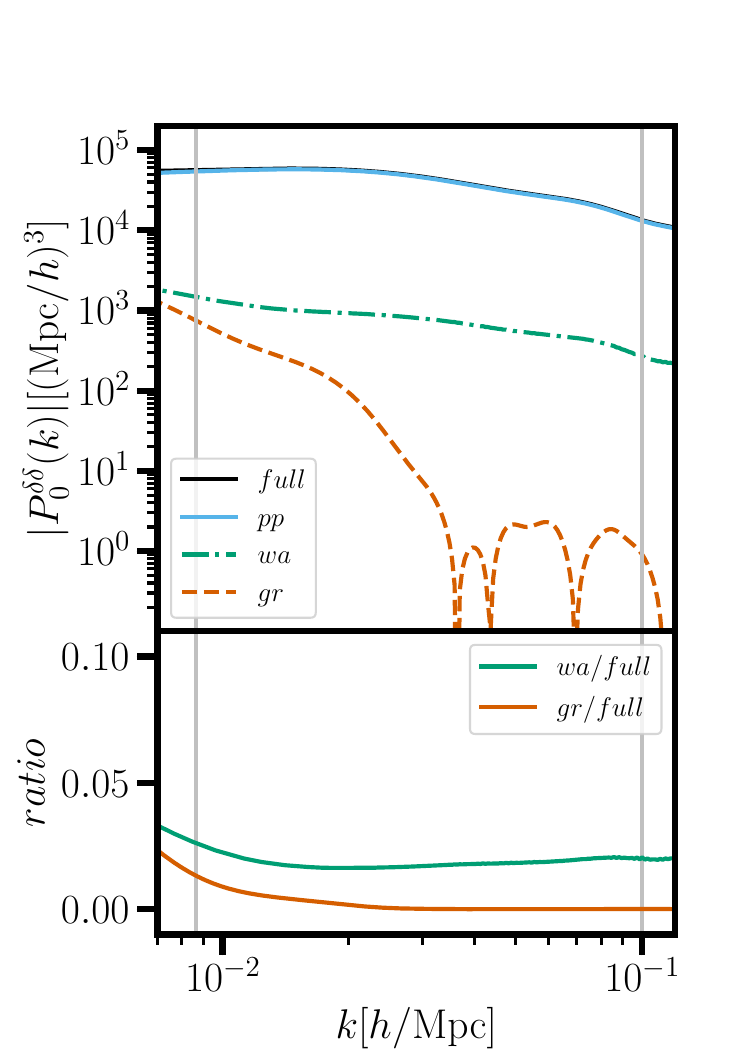}
\includegraphics[width=0.32\textwidth]{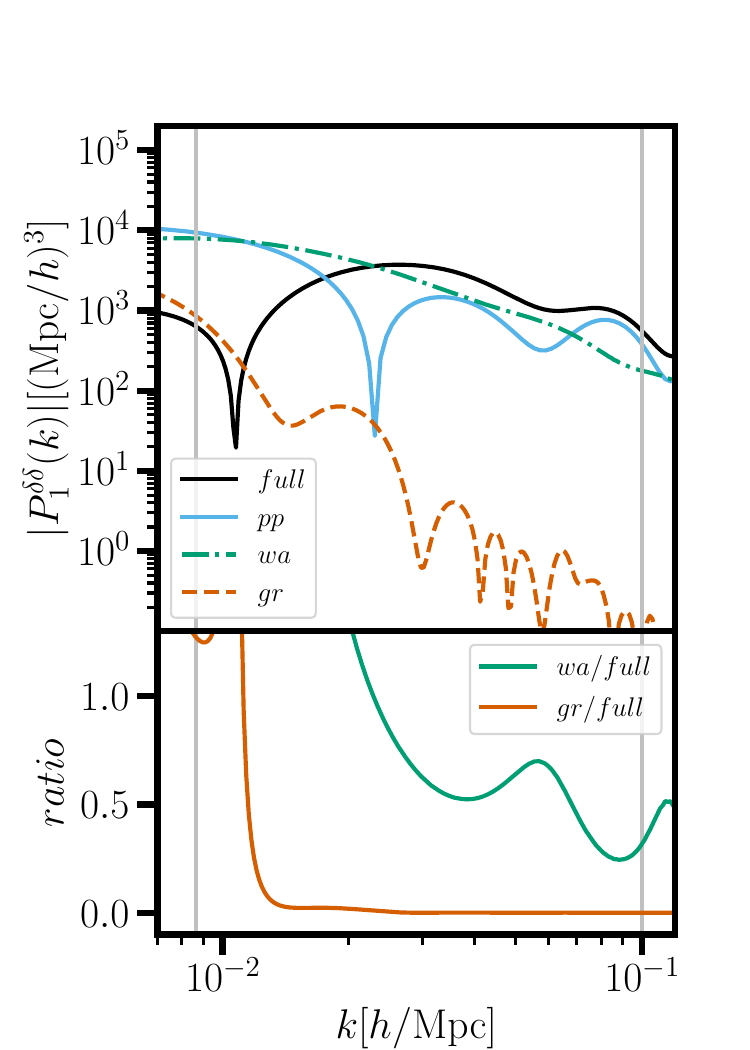}
\includegraphics[width=0.32\textwidth]{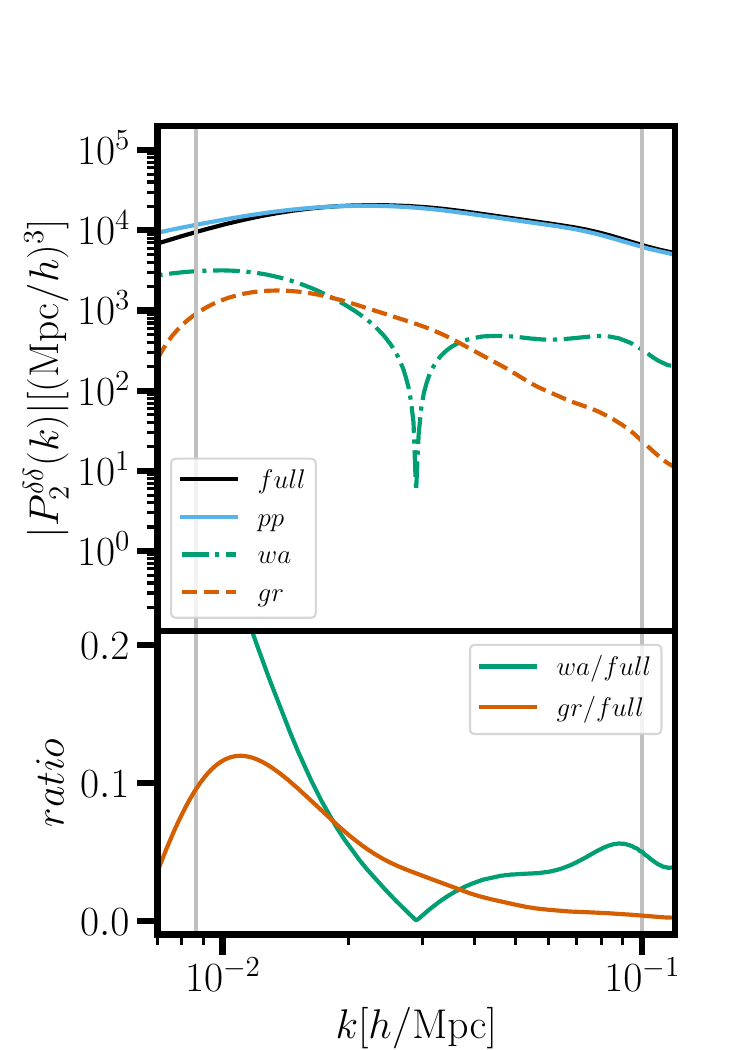}
\includegraphics[width=0.32\textwidth]{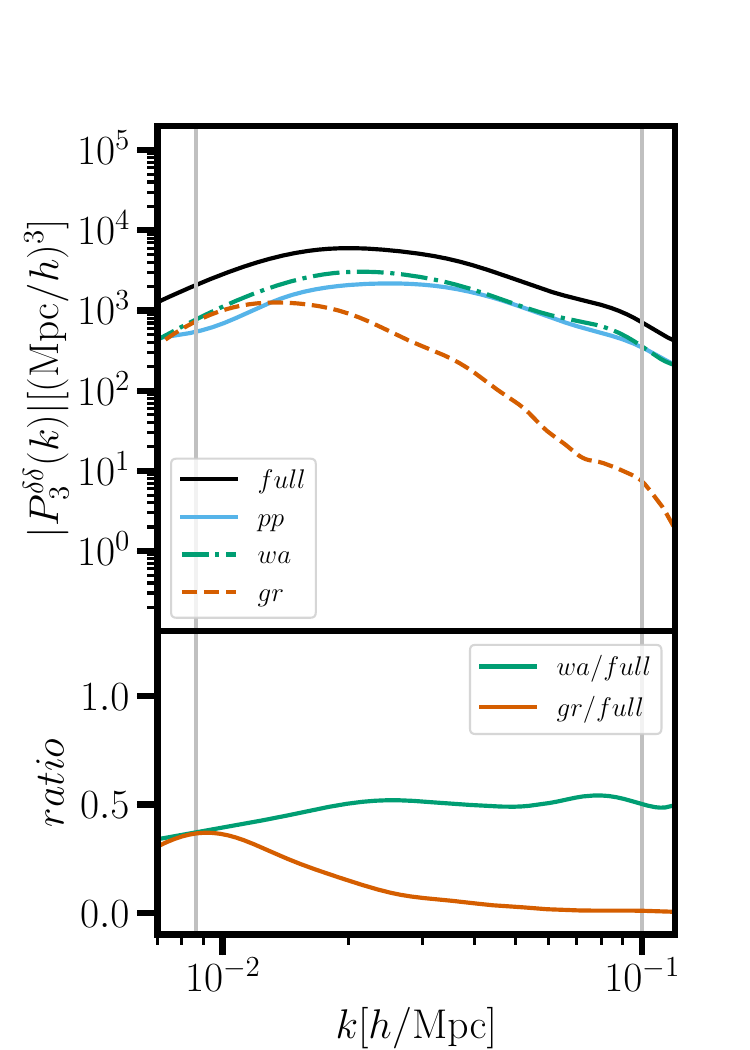}
\includegraphics[width=0.32\textwidth]{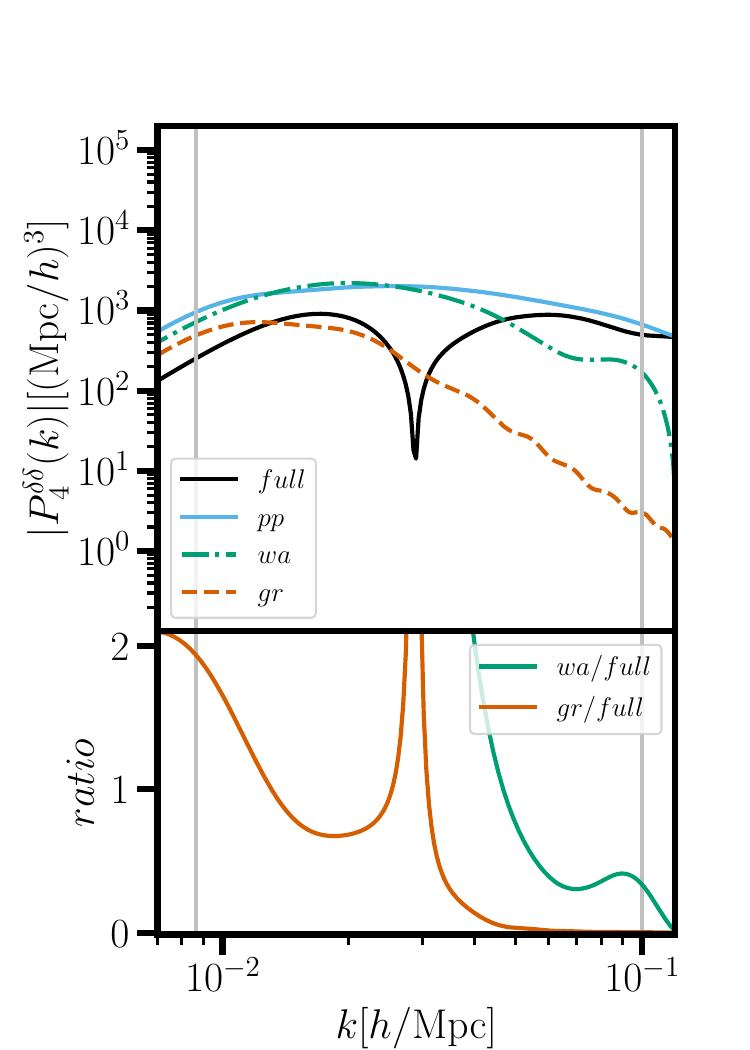}
    \caption{The multipoles of the galaxy number counts power spectrum $P^{(\delta \delta)}_L(k)$ up to $L=4$. We consider a survey covering half of the sky between $z_{\rm min} = 0.05$ $z_{\rm max} = 0.2$ with a constant number density. In all panels the full theory expressions is shown with continuous black lines, the plane-parallel limit in cyan, the wide-angle term with dot-dashed green lines, and the GR term in orange dashed lines. In each plot, the lower panels displays the ratio between wide-angle or GR terms and the full expression for the multipoles.}
    \label{fig:Pdd}
\end{figure}

For completeness, in this appendix we show, in Fig.~\ref{fig:Pdd}, the multipoles of the galaxy number counts power spectrum used in the Fisher matrix analysis of Sec.~\ref{sec:forecasts}. For the monopole, displayed in the upper-left panel, both wide-angle and GR corrections are smaller than 5\% of the total power over all relevant scales. In the quadrupole, wide-angle and GR terms can reach, at low-$k$, 30\% and 10\% of the total signal respectively. The hexadecapole, shown in the bottom-right panel, is severely affected by wide-angle corrections, which have the same size of the plane-parallel piece but opposite sign, leading to large cancellations for $k\lesssim 0.05\, h/$Mpc. The relativistic contributions therefore dominate the signal over these range of scales. Finally, the dipole and octopule are mostly sourced by wide-angle terms and leakages from the even multipoles introduced by the window function.

\bibliographystyle{JHEP}
\bibliography{main}

\end{document}